\documentclass[aps,prd,twocolumn,preprintnumbers,nofootinbib,showpacs,floatfix]{revtex4}

\usepackage{verbatim}
\usepackage{graphicx}
\usepackage{dcolumn}
\usepackage{bm}

\newcommand{\mathscr}[1]{{\cal #1}}

\newcommand{\etal}{{\it et al.}}
\newcommand{\tmin}{t_+\left(1-\sqrt{1-t_-/t_+}\right)}

\begin{document}

\preprint{CLNS 06/1968}       
\preprint{CLEO 06-13}         

\title{A Study of the Semileptonic Charm Decays 
  $D^0 \to \pi^{-} \lowercase{e}^{+} \nu_e$, 
  $D^{+} \to \pi^0 \lowercase{e}^{+} \nu_e$, 
  $D^0 \to K^{-} \lowercase{e}^{+} \nu_e$, and 
  $D^{+} \to \bar{K}^0 \lowercase{e}^{+} \nu_e$}

\author{S.~Dobbs}
\author{Z.~Metreveli}
\author{K.~K.~Seth}
\author{A.~Tomaradze}
\affiliation{Northwestern University, Evanston, Illinois 60208}
\author{J.~Ernst}
\affiliation{State University of New York at Albany, Albany, New York 12222}
\author{H.~Severini}
\affiliation{University of Oklahoma, Norman, Oklahoma 73019}
\author{S.~A.~Dytman}
\author{W.~Love}
\author{V.~Savinov}
\affiliation{University of Pittsburgh, Pittsburgh, Pennsylvania 15260}
\author{O.~Aquines}
\author{Z.~Li}
\author{A.~Lopez}
\author{S.~Mehrabyan}
\author{H.~Mendez}
\author{J.~Ramirez}
\affiliation{University of Puerto Rico, Mayaguez, Puerto Rico 00681}
% ...
\author{(CLEO Collaboration)} 
\noaffiliation

\date{December 5, 2007}

\begin{abstract} 
Using a sample of 1.8 million $D\bar{D}$ mesons collected at the $\psi(3770)$ with
the CLEO-c detector, we study the semileptonic decays 
$D^0 \to \pi^- e^+ \nu_e$, 
$D^{+} \to \pi^0 e^+ \nu_e$, 
$D^0 \to K^- e^+ \nu_e$, and
$D^{+} \to \bar{K}^0 e^+ \nu_e$. For the total  branching fractions we 
find 
${\cal B}(D^0 \to \pi^- e^+ \nu_e) = 0.299(11)(9)\%$, 
${\cal B}(D^+ \to \pi^0 e^+ \nu_e) = 0.373(22)(13) \%$,
${\cal B}(D^0 \to K^- e^+ \nu_e) = 3.56(3)(9) \%$, and 
${\cal B}(D^+ \to \bar{K}^0 e^+ \nu_e) = 8.53(13)(23) \%$,
where the first error is statistical and the second systematic. 
In addition, form factors are studied through fits to the partial branching fractions obtained
in five $q^2$ ranges.  By combining our results with recent unquenched lattice calculations,
we obtain $|V_{cd}| = 0.217(9)(4)(23)$ and 
$|V_{cs}| = 1.015(10)(11)(106)$, where the final error is theoretical. 
\end{abstract}

\maketitle

\section{Introduction}
\label{sec:intro}
In the standard model of particle physics, mixing of the quark mass 
eigenstates in their charged current interactions is
described by the Cabibbo Kobayashi Maskawa (CKM)
matrix~\cite{km}. This $3 \times 3$ quark mixing matrix must
be unitary and can be described by four independent parameters. If the
standard model is complete, experimental determination of the CKM matrix
elements should verify its unitarity. Deviations from unitarity would
indicate the presence of physics beyond the standard model. A variety of 
$CP$-conserving and $CP$-violating
observables probe the elements of the CKM matrix and allow us to over-constrain
it. Many of the key observables require great
precision or great sensitivity to provide the constraints at the level needed to
test the validity of the standard model description. It thus remains a continuing
experimental challenge to test the unitarity of the CKM matrix fully.  

Study of the semileptonic decay of $D$ mesons plays a primary role in
our understanding of the CKM matrix. These decays allow
robust determination of the CKM matrix elements $|V_{cs}|$ and $|V_{cd}|$ by combining
measured branching fractions with form factor calculations, such as those based on unquenched lattice
QCD (LQCD) \cite{fnalqcd}. In addition, these measurements will provide precision tests of
LQCD itself~\cite{yellowbook}. One approach to tests of LQCD assumes
unitarity of the CKM matrix and compares the constrained matrix elements
\cite{pdg} to elements obtained with a combination of CLEO-c measurements and
lattice form factors.  A second approach, which is independent of CKM elements
and thus free from the unitarity assumption, 
compares the measured and
calculated ratios of semileptonic and purely leptonic branching
fractions.
Verification of lattice calculations at the few percent level will 
provide validation for use of the lattice in the $B$ system,
where they are relied upon for several crucial theoretical
quantities.  

This article presents a study of the 
$D^0 \to \pi^- e^+ \nu_e$, 
$D^{+} \to \pi^0 e^+ \nu_e$,
$D^0 \to K^- e^+ \nu_e$ and
$D^{+} \to \bar{K}^0 e^+ \nu_e$
decay modes (charge conjugate modes
implied). A summary of the analysis is also provided in a shorter 
companion article~\cite{PRL}. The results are based on a sample of $1.8$
million $D\bar{D}$ pairs
collected with the CLEO-c detector at the Cornell Electron Storage Ring (CESR) 
from 281 pb$^{-1}$ of $e^+e^-$ data at the $\psi (3770)$ resonance. The sample is a
superset of, and approximately five times larger than, the data used to obtain the first
CLEO-c semileptonic branching fraction measurements~\cite{dsemilep}.
For each mode we determine the partial branching fractions in
five $q^2$ ranges, with the sum of the five rates determining the
total branching fraction. Fits to the rates determine
the form factor shapes. By incorporating LQCD calculations into 
the form factor fits, we extract values for the CKM elements $|V_{cd}|$ 
and $|V_{cs}|$. Previous quenched lattice predictions 
carried errors of about $20\%$. Current unquenched LQCD
calculations allow theoretical
evaluation of the form factors at the $10\%$~\cite{fnalqcd} level,
with future improvement to the few percent level expected.

Within this article, Section~\ref{sec:semilep} provides an overview of
the formalism for exclusive semileptonic decays of charm mesons and their
associated form factors. Sections~\ref{sec:event_recon} through~\ref{sec:bf_results} cover the
experimental procedures for event reconstruction and extraction of the
branching fractions, the systematic uncertainty evaluation, and the
branching fraction results.  Sections~\ref{sec:ff} and~\ref{sec:ckm} explore the form
factor shape constraints from our data and the extraction of $|V_{cs}|$ and
$|V_{cd}|$. Section~\ref{sec:summary} presents our conclusions and comparisons with
previous measurements.

\section{Exclusive Charmed Semileptonic Decays}
\label{sec:semilep}
The matrix element describing the semileptonic decay of a $D$ meson to a pseudoscalar
meson $P$ is of the form
\begin{equation}
\label{sl_amp}
\mathscr{M}(D \to P e^+ \nu_e) = 
-i\frac{G_F}{\sqrt{2}}V_{cq}^*L^{\mu}H_{\mu},
\end{equation}
where $G_F$ is the Fermi constant, $V_{cq}$ is the appropriate CKM matrix
element and $L^\mu$ and $H_\mu$ are the leptonic and hadronic
currents. The leptonic current can be written in terms of the electron
and neutrino Dirac spinors, $u_e$ and $v_\nu$,
\begin{equation}
L^\mu = \bar{u}_e\gamma^\mu(1-\gamma_5)v_\nu.
\end{equation}
In the case of pseudoscalar decays, where there is no axial-vector
contribution, the hadronic current is given by
\begin{equation}
H_\mu = \left< P(p) | \bar{q}\gamma_\mu c | D(p') \right>,
\end{equation}
where $p'$ and $p$ are the four-momenta of the parent $D$ meson and the 
daughter $P$ meson, respectively.  The hadronic current is fundamentally a
non-perturbative quantity that is difficult to evaluate.  We can, however, re-parameterize
the current by expressing it in terms of the independent four-momenta in the
process, which for a pseudoscalar-to-pseudoscalar decay are the two
four-momenta $p' + p$ and $q = p' - p$. We can identify $q$ as the four-momentum of the virtual $W$
boson. A typical formulation of the hadronic current in terms of these four-momenta is given by
\begin{eqnarray}
\lefteqn{\left< P(p) | \bar{q}\gamma^\mu c | D(p') \right>  = } \\ \nonumber
 & &f_+(q^2)\left[ (p' + p)^\mu -\frac{M_D^2-m_P^2}{q^2}q^\mu\right] + 
 \\ \nonumber 
 & & f_0(q^2)\frac{M_D^2-m_P^2}{q^2}q^\mu,
\end{eqnarray}
where $M_D$ is the mass of the $D$ meson and $m_P$ is the mass of the
final state pseudoscalar meson. 
The non-perturbative contributions are incorporated in the scalar functions
 $f_+(q^2)$ and $f_0(q^2)$, the form factors of the decay.  Kinematic constraints require $f_+(0) = f_0(0)$.
 A further simplification arises due to the small mass of the
electron because $q^\mu L_\mu \to 0$ in the limit $m_e \to 0$. Thus including only the $f_{+}$ form factor in the hadronic current,
\begin{equation}
\label{finalhadc}
\left< P(p) | \bar{q}\gamma^\mu c | D(p') \right> = 
f_+(q^2)(p' + p)^\mu,
\end{equation}
is a very good approximation.
With this form for the hadronic current the partial decay
width becomes
\begin{equation}
\label{slwidth}
\frac{d\Gamma(D \to Pe\nu_e)}{dq^2} = 
\frac{G_F^2|V_{cq}|^2}{24\pi^3}p^3|f_{+}(q^2)|^2.
\end{equation}

The partial decay width (Eq.~\ref{slwidth}) clearly reveals that extraction of the CKM matrix elements from measured rates requires prediction of the semileptonic form factors. Theoretical calculation of the form factors therefore has become a
considerable industry.   We focus here on parameterizations of the form factors
that we employ in our form factor studies and in extraction of $|V_{cd}|$ and $|V_{cs}|$.

The goal of any particular parameterization of the semileptonic form
factors is to provide an accurate, and physically meaningful, expression
of the strong dynamics in the decays. To that end, one may
express the form factors in terms of a dispersion
relation, an approach that has been well established in the literature 
(see for example Ref.~\cite{boyd_95} and references therein). It is common to
write the dispersive representation in terms of an explicit pole and a sum of effective poles:
\begin{equation}
\label{disp_effp}
f_+(q^2) = \frac{f_+(0)}{1-\alpha}\frac{1}{1-\frac{q^2}
  {m_{\mathrm{pole}}^2}} +
  \sum_{k=1}^{N}\frac{\rho_k}{1-\frac{1}{\gamma_k}\frac{q^2}{m_{\mathrm{pole}}^2}},
\end{equation}
where $\rho_k$ and $\gamma_k$ are expansion parameters. Given the underlying
$c \to q$ quark transition of the semileptonic
decay, the mass $m_{\mathrm{pole}}$ is the mass of the lowest-lying $c\bar{q}$
vector meson.
The parameter $\alpha$ gives the contribution from the vector meson pole at
$q^2 = 0$. Using this dispersion relation the true form factor can be approximated 
to any desired degree of accuracy by keeping sufficient terms in the
expansion. This approach has the drawback that the decay
dynamics are not explicitly predicted. Additionally, experimental data
have suggested the need for only a few parameters in the description of the
form factor shape. It is therefore natural to seek simplifications of
this parameterization that can still capture the correct dynamics.

Removing the sum over
effective poles entirely, leaving only the explicit vector meson
pole, provides one simplification route  that is typically referred to as ``nearest pole dominance'' 
or ``vector-meson dominance''.  The resulting
``simple pole'' parameterization of the form factor is given by
\begin{equation}
\label{spole}
f_+(q^2) = \frac{f_+(0)}{(1-\frac{q^2}{m^2_{\mathrm{pole}}})}.
\end{equation} 
Experimental data disagree with the physical
basis for this approximation, since measurements of the parameter
$m_{\mathrm{pole}}$ that fit the data do not agree with the expected vector
meson masses~\cite{hill_fpcp_talk_writeup}.
Effectively, at low or medium values of $q^2$ the spectrum is distorted compared to
the simple pole model, receiving contributions from the continuum of effective poles
above the lowest lying pole mass. 

The modified pole, or Becirevic-Kaidalov (BK)
parameterization~\cite{BKparam}, was proposed to address this
problem. The parameterization keeps the first term from the effective pole
expansion, while making simplifications such that the form factor can be
expressed using only two parameters: the intercept 
$f_+(0)$ and an additional shape parameter.\footnote{There will be three 
parameters if the $f_0(q^2)$ form factor, which we are neglecting due to the small
electron mass, is also taken into account.} 
The parameterization is typically expressed in the form
\begin{equation}
\label{mod_pole}
f_+(q^2) = \frac{f_+(0)}{(1-\frac{q^2}{m^2_{\mathrm{pole}}})
(1-\alpha\frac{q^2}{m^2_{\mathrm{pole}}})}.
\end{equation}
This parameterization has recently been widely used in the extraction of semileptonic form
factors from experimental measurements~\cite{cleo_2005,focus_2005,belle_2006,Aubert:2006mc}.
In addition, some recent LQCD calculations of the form factor
have relied on this parameterization for extrapolation and interpolation
purposes~\cite{fnalqcd,hpqcd_vub}. This scheme
requires several assumptions to reduce the
multiple parameters initially present (Eq.~\ref{disp_effp})  to
one. The BK ansatz assumes that the
gluon hard-scattering contributions ($\delta$) are close to zero and that scaling
violations ($\beta$) are close to unity, which may be succinctly expressed as
\begin{equation}
1 + 1\slash \beta - \delta \equiv 
\frac{\left(M_D^2 - m_{P}^2\right)}
{f_+(0)} \left.\frac{df_+}{dq^2}\right|_{q^2=0} \sim 2.
\end{equation}
Once again, however, the experimental data do not bear out 
these assumptions~\cite{hill_fpcp_talk_writeup}. We should observe $\alpha \sim 1.75$ in order
to obtain $1 + 1\slash \beta - \delta = 2$, whereas the observed data are removed from
such values by many standard deviations. 

We note that both functional forms can provide adequate parameterizations of
the data if their parameters are allowed to be non-physical.  Without a physical
underpinning for the parameterization, however, parameters obtained
from theory and/or from different experiments may not agree if
their form factor sensitivities differ as a function of $q^2$.

Our primary form factor shape analysis therefore utilizes a series
expansion around $q^2=t_0$ that has been advocated by several groups for a
physical description of heavy meson form factors~\cite{boyd_95,boyd_97,grinstein,rhill}.
The series expansion is congruous with the dispersion relations, 
and is guaranteed to contain the true form factor, yet is still
rich enough to describe all
variations that affect the physical observables. 

To achieve a convergent series, the expansion is formulated as an analytic 
continuation of the form factor into the complex $t=q^2$ plane. There is
a branch cut on the real axis for $t>(M_D+M_{K,\pi})^2$ that is mapped
onto the unit circle by the variable $z$, defined as
\begin{equation}
z(q^2,t_0) = \frac{\sqrt{t_+ - q^2} - \sqrt{t_+ - t_0}}{\sqrt{t_+ - q^2}
	  + \sqrt{t_+ - t_0}}, 
\end{equation}
where $t_\pm \equiv (M_D \pm m_{K,\pi})^2$ and $t_0$ is the (arbitrary) $q^2$ value
that maps to $z=0$.  The expression for the form factor becomes
\begin{equation}
\label{eq:zexpand}
f_+(q^2) = \frac{1}{P(q^2)\phi(q^2,t_0)}\sum_{k=0}^\infty
a_k(t_0)[z(q^2,t_0)]^k, 
\end{equation}
with
\begin{equation}
P(q^2) \equiv \left\{ \begin{array}{rl} 1, & D \to \pi \\
z(q^2,M^2_{D^*_s}), & D \to K \\
\end{array} \right. .
\end{equation}
The $P(q^2)$ factor accommodates sub-threshold resonances, which overcomes
the convergence issues that a naive expansion would face with a nearby pole. 
Good convergence properties are expected since
the physical region is restricted to $|z|<1$.
The physical observables do not depend on the choice of 
$\phi(q^2,t_0)$, which can be any analytic function, or on the value of
$t_0$. We report $a_k$ parameters that correspond to $t_0 = 0$ and the ``standard'' 
choice for $\phi$ (see, {\it e.g.} Ref.~\cite{rhill} and Appendix~\ref{app:zexpand}), which
results from bounding $\sum a_k^2$ from unitarity considerations. Appendix~\ref{app:zexpand}
presents results for an alternate choice of $t_0$ that minimizes the maximum
value of $|z|$ over the physical range.
If the series converges quickly, as expected, it is
likely that only the first two or three terms will be able to be seen in
the data. We will explore the number of terms needed to adequately
describe our data.

While our primary form factor and CKM results will be based on the series expansion, 
we will also provide results based on the two
pole parameterizations for comparative purposes.
   
\section{Event Reconstruction and Selection}
\label{sec:event_recon}
The analysis technique rests upon association of the missing energy and momentum 
in an event with the neutrino four-momentum \cite{nu_recon_other}, an
approach enabled by the excellent hermeticity and resolution of the CLEO-c detector 
\cite{yellowbook,cleoiii_and_zd_nims}. 
Charged particles are detected over $93\%$ of the solid angle by two wire
tracking chambers within a $1.0$~T solenoid magnet. The momentum
resolution is $0.6\%$ at $800$ MeV/$c$. 
Specific ionization measurements from the tracking system in combination with a ring imaging \v Cerenkov detector (RICH) \cite{rich_nim_or_equiv} provide particle identification. A CsI(Tl) crystal
electromagnetic calorimeter provides coverage over about $93\%$ of $4\pi$, and achieves
a typical $\pi^0$ mass resolution of $6$ MeV/$c^2$. 

Electron candidates are identified above $200$ MeV$/c$  over
$90\%$ of the solid angle by combining
information from specific ionization with calorimetric, RICH and
tracking measurements. 
The identification efficiency, which has been determined from data, is 
greater than $96\%$ above $500$ MeV$/c$  and
greater than $90\%$ above $300$ MeV$/c$. The
average probability that a hadron is misidentified as an electron is less than
$0.8\%$.  Below $300$ MeV$/c$ the efficiency falls rapidly, reaching 60\% in the $200-250$
MeV$/c$ region. To reduce our sensitivity to final state radiation (FSR), we add
photons within 3.5$^\circ$ of the initial electron momentum
back into the tracking-based four-momentum.

Charged pions and kaons from the signal decay are identified using
specific ionization and RICH
measurements. Pion candidates below 750 MeV$/c$ and kaon candidates
below 500 MeV$/c$ are identified using only specific ionization information,
which is required to be within three standard deviations ($\sigma$) of
that expected for the assigned particle type. For pion
candidates above 650 MeV$/c$, we also require the pion mass hypothesis
be more likely than the kaon mass hypothesis. Above these momenta, candidate tracks must also pass
RICH identification criteria. Specifically, we require that pion (kaon) candidates are
more than 3$\sigma$ closer to a pion (kaon) hypothesis than a kaon (pion) hypothesis. 

A $\pi^0$ candidate must have a $\gamma\gamma$ mass within
2.5$\sigma$ of the $\pi^0$ mass. $K_S^0$ candidates are
reconstructed using a constrained vertex fit to candidate $\pi^+\pi^-$
daughter tracks. The $\pi^+\pi^-$ mass must  be within 4.5$\sigma$ of the $K_S^0$ mass.

To reconstruct the undetected neutrino we utilize the hermeticity of the CLEO-c
detector to find the missing energy and momentum. In the process $e^+ e^-
\to \psi(3770) \to D\bar{D}$, the total energy of the beams is imparted to
the $D\bar{D}$ system. Because the beam energies
are symmetric and the beam crossing angle is small at CESR, each produced
$D$ has an energy of the beam energy to within a small correction.
The missing four-momentum in an event is given by $p_{\mathrm{miss}} =
(E_{\mathrm{miss}},\vec{p}_{\mathrm{miss}})= 
 p_{\mathrm{total}} - \sum p_{\mathrm{charged}} - \sum
 p_{\mathrm{neutral}}$, where the event
four-momentum $p_{\mathrm{total}}$  is known from the energy
and crossing angle of the CESR beams. Charged
and neutral particles for the sums must pass selection criteria designed
to achieve the best possible $|\vec{p}_{\mathrm{miss}}|$ resolution by
balancing the efficiency for detecting true particles against the
rejection of false ones.

For the charged four-momentum sum, $\sum
p_{\mathrm{charged}}$, optimal selection is achieved with topological criteria.
These criteria minimize
multiple-counting that can result from low-momentum tracks that curl in the
magnetic field, charged particles that decay in flight or interact within
the detector, and spurious tracks. Tracks that are actually segments of a single low transverse
momentum ``curling'' particle are identified by selecting reconstructed
track pairs with opposite curvature whose innermost and outermost diametric radii each match
within 14 cm and whose separation in $\phi$ is within
180$^\circ \pm 20^{\circ}$.  For physics use we select the track segment that
will best represent the original charged particle based on track
quality and distance-of-closest-approach to the beam spot.
We employ similar algorithms to identify particles that
curl more than once, creating three or more track segments.
We also identify tracks that have scattered or decayed in the drift
chamber, causing the original track to end and one or more tracks to
begin in a new direction.  We keep only the track segment with the majority of its hits
before the interaction point.  Spurious tracks are identified by their low hit
density and/or low number of overall hits and rejected.

Each hadronic track must be assigned a mass hypothesis to calculate
its contribution to the total energy sum. We assign a most probable
mass hypothesis by combining detector measurement with particle production
information.   
The production information is introduced because it
is only statistically advantageous to identify a track as a kaon
at a momentum where many more pions than kaons are produced 
when the detector's particle ID information strongly favors a kaon.
For each track,  we
first calculate a likelihood for the kaon and pion hypothesis based on
specific ionization and RICH measurements.  Those likelihoods
are then weighted by the  Monte Carlo (MC) prediction for the relative
$K^-$ and $\pi^-$ abundances in $D$ decays at that track's momentum,
which then gives us the true probability for each mass hypothesis. 

For the neutral four-momentum  sum, $\sum  p_{\mathrm{neutral}}$, clusters
resulting from the interactions of charged hadrons must be avoided.  As a
first step, calorimeter showers passing the standard CLEO
proximity-matching (within 15 cm of a charged track) are eliminated.
Optimizations also revealed that all showers under 50 MeV should be
eliminated.  The processes that result in reconstructed
showers (``splitoffs'') separate from but within about 25$^\circ$ of a proximity-matched
shower  tend to result in an energy distribution over the $3\times3$
central array of the splitoff shower that ``points back'' to the core
hadronic shower.  We combine this information with the ratio of
energies in the $3\times3$ to $5\times5$ arrays of crystals, whether the
shower forms a good $\pi^0$, and the MC predictions for relative
spectra for true photons versus splitoff showers to provide an optimal
suppression of the contribution.

Association of the missing four-momentum with the neutrino
four-momentum is only accurate if the event contains no more than one
neutrino and if all true particles are detected.  For events with
additional missing particles or doubly-counted particles, the signal
modes tend not to reconstruct properly while background processes tend
to smear into our sensitive regions.  Hence, it is worthwhile to reject
events for which independent measures indicate these problems.  
We therefore exclude events that have either more than one electron or non-zero
net charge.  Multiple electrons indicate an increased likelihood for
multiple neutrinos, while non-zero net charge indicates at least one
missed or doubly-counted charged particle.

After application of the above criteria approximately 90\% of the signal MC
$|\vec{p}_{\mathrm{miss}}|$ distribution is contained in a central core with
$\sigma \sim 15$ MeV$/c$.  

To further enhance the association of the missing momentum with an
undetected neutrino in our final event sample, we require that the 
$M_{\mathrm{miss}}^2 \equiv E_{\mathrm{miss}}^2 -
|\vec{p}_{\mathrm{miss}}|^2$ be consistent with 
a massless neutrino.  The $M_{\mathrm{miss}}^2$ resolution,
$$
\sigma(M_{\mathrm{miss}}^2) = 2E_{\mathrm{miss}}\sigma(E_{\mathrm{miss}}) \oplus  2|\vec{p}_{\mathrm{miss}}|\sigma(|\vec{p}_{\mathrm{miss}}|),
$$
is dominated by the $E_{\mathrm{miss}}$ term since the
resolution of $|\vec{p}_{\mathrm{miss}}|$ is roughly half that
of $E_{\mathrm{miss}}$.  MC simulation indicated an optimal requirement of
$\left|M_{\mathrm{miss}}^2 \slash 2|\vec{p}_{\mathrm{miss}}|\right| < 0.2$ GeV$/c^3$,
which (noting $E_{\mathrm{miss}}\approx|\vec{p}_{\mathrm{miss}}|$ for
signal) provides selection at approximately constant $E_{\mathrm{miss}}$ resolution.
Additionally, because of the superior $|\vec{p}_{\mathrm{miss}}|$ resolution, in
subsequent calculations we take 
$p_{\nu} \equiv (|\vec{p}_{\mathrm{miss}}|,\vec{p}_{\mathrm{miss}})$.

Semileptonic decays $D \to P e \nu$, where $P$ is a pion or kaon, are
identified by their consistency with the expected $D$ energy and
momentum. Candidates are selected based
on $\Delta E \equiv (E_P + E_e + E_\nu) - E_{\mathrm{beam}}$ (expected to
be zero within our resolution of about $20$ MeV) and yields are
extracted from the resulting distributions in beam-constrained mass
$M_{\mathrm{bc}}$ (equivalent to $D$ momentum and expected to be close to the
known $D$ mass). These quantities are corrected for the small boost resulting
from the 3 mrad beam crossing angle. Because the $|\vec{p}_\nu|$ resolution dominates the
$\Delta E$ resolution, we can improve our $p_\nu$ measurement by scaling it 
by the factor $\zeta$ satisfying $(E_P + E_e + \zeta E_\nu) - E_{\mathrm{beam}}=0$.
We use $\zeta \vec{p}_\nu$ for the neutrino momentum in
computation of $M_{\mathrm{bc}} \equiv \sqrt{E_{\mathrm{beam}}^2 -
  |\vec{p}_{P} + \vec{p}_{e} + \zeta\vec{p}_\nu|^2}$. The resulting resolution
for $M_{\mathrm{bc}}$ is 4 MeV$/c^2$.

Selection criteria were optimized by studying MC samples independent of
those used elsewhere in the analysis. Sources of
backgrounds include events with fake electrons, non-charm continuum
production ($e^+e^- \to q\bar{q}$, $e^+e^- \to \tau^+\tau^-$ and 
$e^+e^- \to \gamma\psi(2S)$), and $D\bar{D}$ processes other than signal.

The optimal $\Delta E$ requirement was determined to be $-0.06 < \Delta
E < 0.10$ GeV. For the Cabibbo-favored modes, the background level remaining
after this selection is only a few percent of the signal level.  
For the Cabibbo-suppressed modes, there remains significant background from
cross-feed among the signal modes, particularly from the kaon modes, as well as from
the related modes $D^+ \to K_L^0 e^+ \nu_e$ and $D^+ \to K_S^0 e^+ \nu_e$ where 
$K_S^0\not\rightarrow \pi^+\pi^-$. 
Since the cross-feed typically involves particles from the ``other $D$'' decay, we obtain
some suppression of this background with a $q^2$--dependent requirement on  $\Delta E_{\text{n.s.}}$ 
for the non-signal particles in the event.  
We obtain $\Delta E_{\text{n.s.}}$ by summing the energy of all non-signal particles in an
event, even though we do not
specifically reconstruct the non-signal $D$ decay.
This criterion effectively imposes an additional
constraint on the quality of the reconstructed neutrino.  
We also require 
$D^+ \to \pi^0 e^+ \nu_e$
candidates to have the smallest $|\Delta E|$ compared to any other
final state candidates in the event, and that these events contain no 
reconstructed $D^0 \to K^- e^+ \nu_e$ candidate. These criteria suppress cross-feed from the charged pion and kaon modes with almost no loss of true $\pi^0 e^+ \nu_e$ decays.
 The average background 
level ($q^2$--dependent) in the pion modes is about 20\% of
the signal level.

To simplify the statistical interpretation of our
results,  as well as to suppress cross-feed from the Cabibbo-favored into
the Cabibbo-suppressed modes, we limit the
number of multiple entries per event  
such that a given event can contribute to at most one $D^0$ or one $D^+$ final state.
For events with multiple $D^+$ candidates or multiple $D^0$ candidates
satisfying $M_{\mathrm{bc}} > 1.794$ GeV$/c^2$, we choose 
the candidate with the smallest $|\Delta E|$, independent of $q^2$. 

From the measured electron and the re-scaled
neutrino four-momenta we calculate $q^2 \equiv M_{W^*}^2$ from $q^2 = (p_{\nu}
+ p_e)^2$. The resulting resolution is 0.01 GeV$^2/c^4$, independent of
$q^2$.

\section{Extraction of Branching Fractions}
\subsection{Method and Binning}
For each of the four signal modes we construct the $M_{\text{bc}}$ distributions
in five $q^2$ ranges: $q^2 < 0.4$ GeV$^2/c^4$, $0.4 \le q^2 < 0.8$
GeV$^2/c^4$, $0.8 \le q^2 < 1.2$ GeV$^2/c^4$, $1.2 \le q^2 < 1.6$
GeV$^2/c^4$ and $q^2 \ge 1.6$ GeV$^2/c^4$. These 20 distributions are
fit simultaneously to extract the partial branching fraction for each interval.
The total branching fraction is then obtained in each mode by summing its five
partial branching fractions.
Fitting in five $q^2$ ranges minimizes the experimental sensitivity of the total
branching fractions to form factor
shape uncertainties, while simultaneous
fitting of all four modes ensures self-consistent handling of the cross-feed
backgrounds among the modes.

The fit utilizes a binned
maximum likelihood approach extended to include the finite statistics
of the MC samples following the method of Barlow and
Beeston~\cite{bbfit}.  The $M_{\mathrm{bc}}$ distribution is divided into fourteen
uniform bins  over the range $1.794 < M_{\mathrm{bc}} <
1.878$ GeV$/c^2$.  

\subsection{Fit Components and Parameters}
\label{sec:fit_comp}
We fit the data to the signal components and five background components.  The signal mode components
are obtained from MC generated using EvtGen \cite{EvtGen} and modified pole-model
form factors~\cite{BKparam} with parameters from the most recent LQCD results~\cite{fnalqcd}. We apply several corrections to our GEANT-based \cite{GEANT}
MC samples to improve simulation of the neutrino reconstruction procedure. 

From independent 
studies, mostly based on CLEO-c samples
with one fully reconstructed hadronic $D$ decay, we evaluate
corrections and associated systematic uncertainties for 
simulation of hadronic showers, false charged
particles and charged-particle identification. We find that the
simulations of charged particles, charged-particle momentum resolution and 
photon-energy resolution need no correction, though we include
the uncertainties in the systematic
uncertainty evaluation. We reweight the MC samples to 
correct the rate and spectrum for $K_L^0$ production (which
affects the neutrino-reconstruction efficiency), for $\pi^0$ and $\pi^-$
production in our full $D$ decay model, and for the momentum-dependent rate 
at which a $K^-$ fakes a $\pi^-$. All of these corrections affect the cross-feed
background rates into and between the Cabibbo-suppressed modes. They
lead to few percent (or less) changes in the measured
yields, but are determined to better than 10\% of themselves.

In the MC samples we select only true electrons (reconstructed tracks that have
been matched to a generator-level electron) with a probability 
for acceptance given by
data-measured efficiencies described earlier. We thereby exclude from the MC any
events caused by identification of a fake electron and
instead estimate this background using data, as we describe in detail below. This
procedure eliminates any reliance on MC predictions for either electron
efficiency or the rate at which hadrons fake electrons. 

We are  sensitive to the distortion of efficiency and kinematics in our
signal modes due to FSR. 
Based on the angular and energy distributions for FSR photons, we correct our 
signal MC, generated with the PHOTOS~\cite{photos} 
package without interference effects included, to the Kaon Leading-Order Radiation (KLOR)~\cite{klor}
calculations modified for charm decay.

For each reconstructed $q^2$ interval in a given mode, we generate a MC
sample in the same (generator level) $q^2$ interval, to which the full
analysis is applied. That is, we obtain the full set of 20 reconstructed
$M_{\text{bc}}$ distributions from each of these 20 independent samples.  
For each of the generated $q^2$ intervals, a single floating parameter, 
which corresponds to the efficiency-corrected data yield in that
interval, controls the normalization of all its 20 reconstructed
distributions. The relative normalizations among
those reconstructed distributions remains fixed at the level predicted by
our corrected MC.  Because the signal rate in each reconstructed range
drives the normalization for the corresponding generated $q^2$ interval,
the data in effect fixes the cross-feed rates into the other 19
reconstructed distributions.

We also use MC samples to describe the $D\bar{D}$ background and the three continuum 
contributions. We absolutely scale the continuum components  
according to their cross
sections at the $\psi(3770)$ and the measured data luminosity. 
The non-signal $D\bar{D}$ sample was generated using EvtGen, with
decay parameters updated to reflect our best knowledge of $D$ meson decays.  
This component floats separately for each reconstructed final state, but
the relative rates over the five $q^2$ regions within that state are fixed.
This approach helps to reduce our sensitivity to
inaccuracies in the $D$ decay model. Finally, we input MC components for 
$D^+ \to K_L^0 e^+ \nu_e$ and $D^+ \to K_S^0(\pi^0\pi^0) e^+ \nu_e$, 
whose rates in each $q^2$
region are tied to those for the signal $D^+ \to K_S^0(\pi^+\pi^-) e^+ \nu_e$ mode.  

The contributions from events in which hadrons have faked
the signal electron are evaluated using data. The momentum-dependent
electron identification fake rates from pions and kaons are measured
using a variety of data samples. We obtain our background estimates
by analyzing a data sample
with no identified electrons. Each track in each event in this sample
is treated in turn as the signal
electron. The contribution in each mode is then weighted according to the fake
rate. The fake electron component is then added to the fit with a fixed,
absolute, normalization.   

Finally, we allow the fit to adjust the
$M_{\mathrm{bc}}$ resolution in the 
$D^0 \to \pi^- e^+ \nu_e$, 
$D^+ \to \pi^0 e^+ \nu_e$, and
$D^0 \to K^- e^+ \nu_e$
modes  by
applying a Gaussian smear to these distributions. As a result the
signal MC $M_{\mathrm{bc}}$ resolution in these modes is increased 
from $\sim$3.5
MeV$/c^2$ to match the data resolution of $\sim$4
MeV$/c^2$. The $M_{\mathrm{bc}}$ resolution in the $D^+ \to K_S^0 e^+ \nu_e$
signal MC matches the data resolution very well 
so we apply no additional smearing to this mode in the fit.

In summary we have 27 free parameters in the fit: the 20 signal rates,
the 4 non-signal $D\bar{D}$ normalizations and the 3 $M_{\mathrm{bc}}$ smearing
parameters. This leaves us with a total of $280 - 27 = 253$ degrees of freedom 
for the fit.  

\subsection{Checks and Results}
\label{sec:fit_candr}

\begin{figure*}
\includegraphics[width=17.8cm]{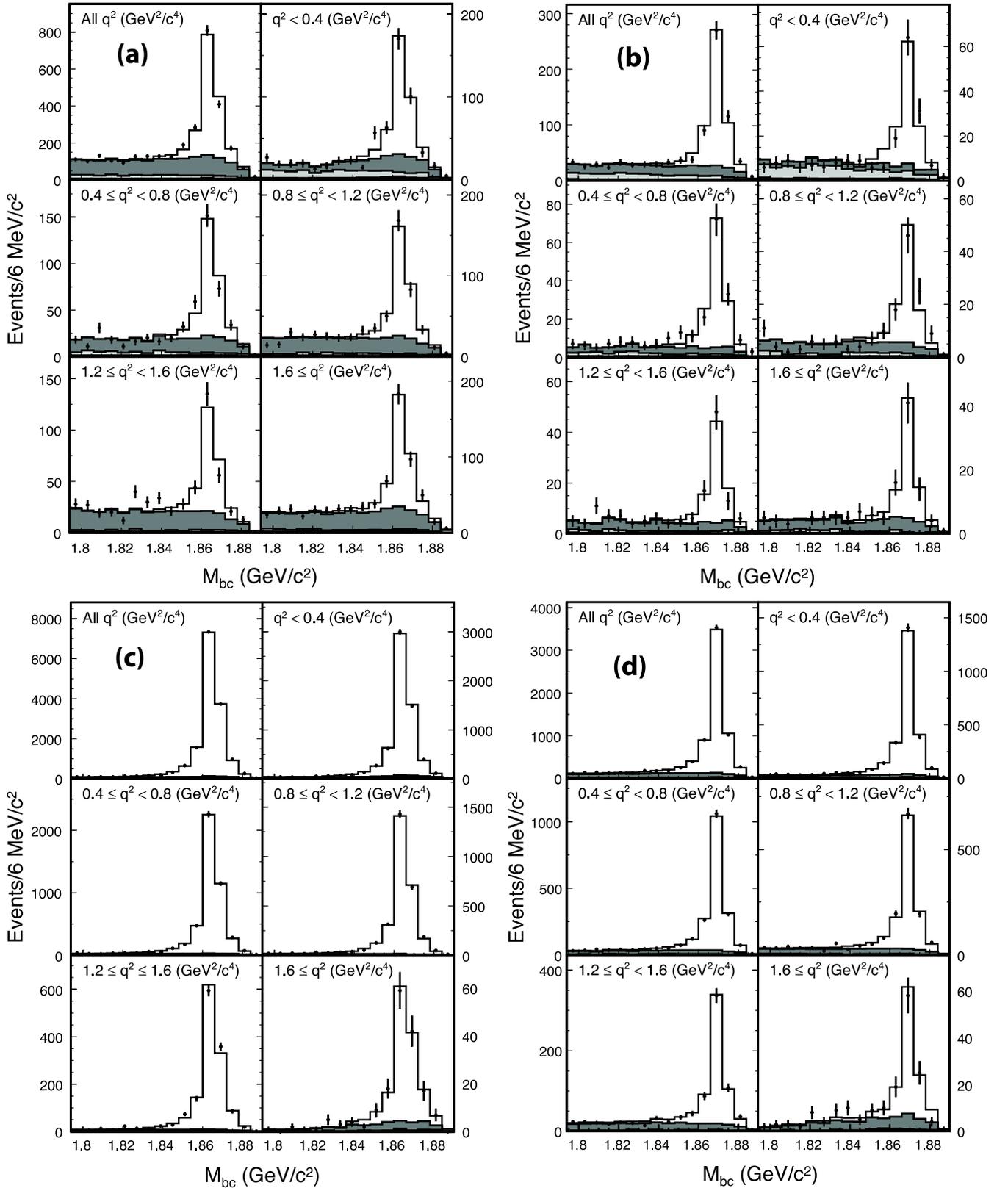} 
\caption{\label{fig:mbc}
 $M_{\mathrm{bc}}$ distributions for the modes (a) $D^0 \to \pi^- e^+ \nu_e$,
(b) $D^+ \to \pi^0 e^+ \nu_e$, (c) $D^0 \to K^- e^+ \nu_e$, and
(d) $D^+ \to K_S^0 e^+ \nu_e$.  The data are shown by the points, and the fit components (histograms) are normalized using the nominal fit results (see text):
signal MC (clear), cross-feed and non-signal $D\bar{D}$ MC
(gray), continuum MC (light gray) and $e^+$ fakes (black).}
\end{figure*}

The direct fit results are displayed as plots in $M_{\mathrm{bc}}$, divided into
the appropriate $q^2$ ranges. These are shown in Fig.~\ref{fig:mbc}.
The value of the likelihood for this fit
is $-2\ln \mathscr{L} = 275.5$ for $253$ degrees of freedom.
Note that each of the 20 distributions
is described by mainly one free parameter -- the signal rate within that
bin, with a more constrained contribution from the parameters
(resolution and $D\bar{D}$ background rate) that float independently for
each signal mode, with the relative contribution into each $q^2$ interval fixed within
a mode.  All other
contributions are either explicitly or effectively fixed by other constraints.

\begin{figure*}
\includegraphics[width=17.8cm]{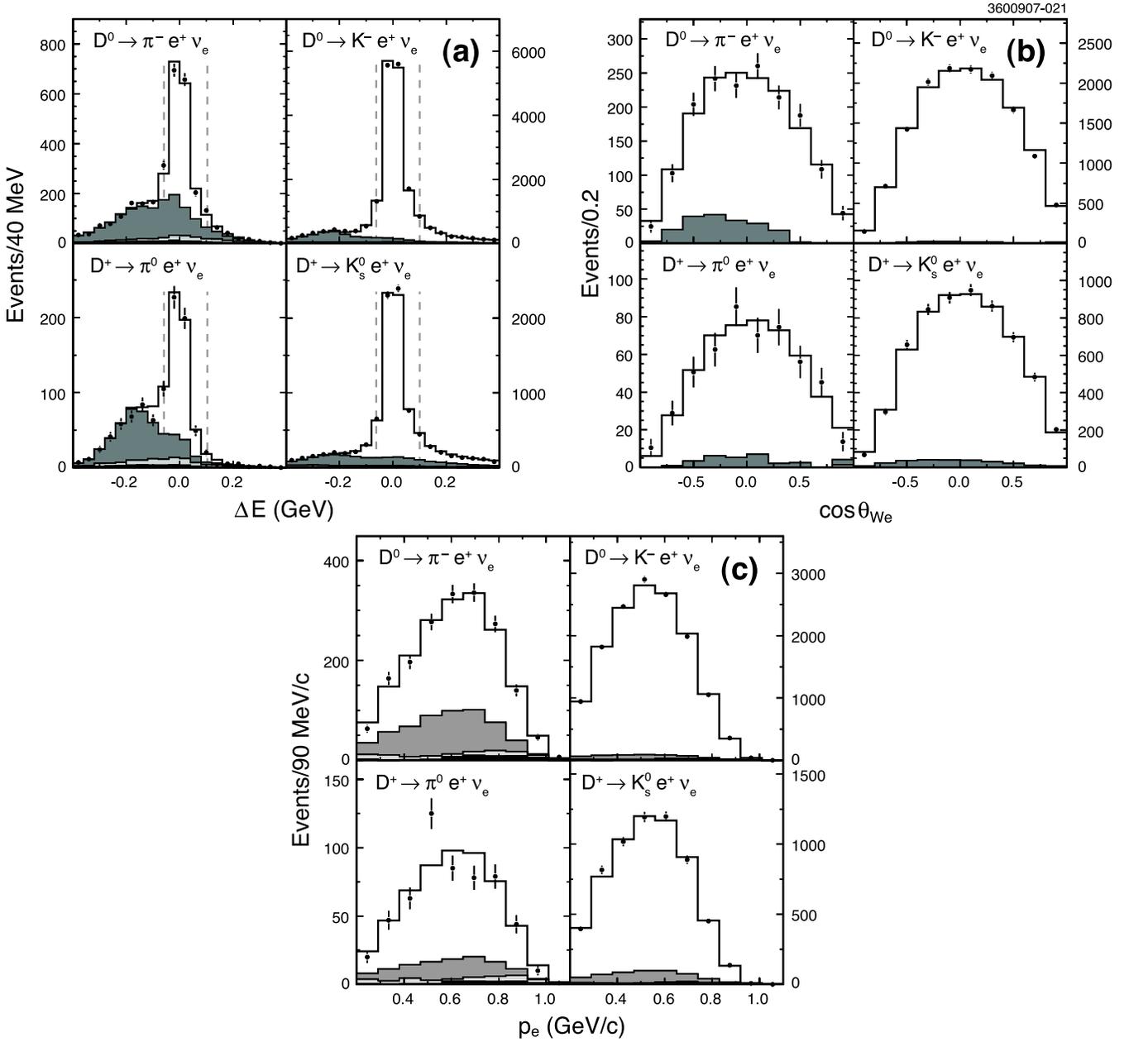} 
\caption{\label{fig:kine} The kinematic distributions for (a) $\Delta E$, (b) $\cos\theta_{We}$, and
(c) $p_{e}$, for events falling within the $M_{\mathrm{bc}}$ signal region for
each of the four signal modes.   The data are shown by the points, and the fit components (histograms) are normalized using the nominal fit results (see text):
signal MC (clear), cross-feed and non-signal $D\bar{D}$ MC
(gray), continuum MC (light gray) and $e^+$ fakes (black).  The dotted lines in (a) indicate for each mode the $\Delta E$ region used in fitting.}
\end{figure*}

Other important reconstructed kinematic variables are
presented integrated over $q^2$ with the components scaled according to
the nominal fit. Fig.~\ref{fig:kine}a shows the $\Delta E$
distributions for events within the signal-enhanced region
$|M_{\mathrm{bc}} - M_D| < 0.015$ GeV$/c^2$. Fig.~\ref{fig:kine}b shows
$\cos\theta_{We}$, the cosine of the angle between the $W$ in the $D$
rest frame and the electron in the $W$ rest frame, in the signal $M_{\text{bc}}$
and $\Delta E$ regions. All of our signal modes should exhibit a
$\sin^2\theta _{We}$ dependence independent of the form factor though
 acceptance effects distort the reconstructed distribution.  The fits
describe the observed distributions very well. 
Finally we find that our fit generally agrees well with the observed momentum ($p_e$) spectrum for the signal electron (Fig.~\ref{fig:kine}c).  The poorest agreement is exhibited by the $\pi^0 e \nu_e$ mode, where the probability of $\chi^2$ is still over 3\%.  
 
To test the fitting procedure, we fit a set of mock data with known input
branching fractions created from the large
$D\bar{D}$ MC sample ($\sim$40 $\times\mathscr{L}_{\mathrm{data}}$) used to
obtain our non-signal
$D\bar{D}$ background estimate. We fit the
sample using distributions from our standard signal MC and from the non-signal portion
of the generic $D\bar{D}$ sample.  Because our ``data'' in this case derives from the same underlying decay model and detector simulation as our fit inputs, we do not apply the corrections noted in the previous section that remove data/MC differences.

\begin{table}[tb]
\caption{ Results of fit to $D\bar{D}$ MC sample with
    statistics of $40\times \mathscr{L}_{\mathrm{data}}$ for all $q^2$
    bins. $Y_{\mathrm{input}} $ is the true
    yield, $Y_{\mathrm{fit}}$ the efficiency corrected yield from the fit, and
    $\sigma_{Y_{\mathrm{fit}}} $ the 1$\sigma$ error on the efficiency-corrected fit yield. }
\label{table:mcfit}
\begin{ruledtabular}
\begin{tabular}{cdddddd}
 & \multicolumn{6}{c}
{$(Y_{\mathrm{input}} - Y_{\mathrm{fit}})\slash\sigma_{Y_{\mathrm{fit}}}$}\\
 & \multicolumn{6}{c}{True $q^2$ interval (GeV$^2/c^4$)}\\
\text{Decay} & \multicolumn{1}{c}{$< 0.4$} & \multicolumn{1}{c}{$0.4-0.8$} 
      & \multicolumn{1}{c}{$0.8-1.2$} & \multicolumn{1}{c}{$1.2-1.6$} 
      & \multicolumn{1}{c}{$\ge 1.6$} & \multicolumn{1}{c}{All $q^2$} \\
\hline
$\pi^- e^+ \nu_e$ &  0.55 & -0.92 & 0.98 &-0.33 & 1.16 & 0.51 \\
$\pi^0 e^+ \nu_e$ & -1.57 &  0.37 &-0.55 & 0.85 & 0.98 &-0.95 \\
$K^- e^+ \nu_e$   & -2.34 &  1.27 & 1.54 & 0.18 & 0.99 &-0.14 \\
$K_S^0 e^+ \nu_e$ & -0.29 & -0.49 & 1.77 & 1.14 & 0.54 & 0.76 \\
\end{tabular}
\end{ruledtabular}
\end{table}

Table~\ref{table:mcfit} presents the
differences between our measured and the generator-level rates.  We see no biases at this
greater level of sensitivity, demonstrating the reliability of the fitting procedure.
Furthermore,
the semileptonic modes in the generic MC sample used to simulate the ``data'' were generated
using ISGW2~\cite{isgw2} form factors, which have a significantly different $q^2$ behavior
than the LQCD-derived form factors~\cite{fnalqcd} of our signal MC. Our test therefore
also verifies that we have adequately subdivided the $q^2$ range to avoid
significant dependence on input form factor modeling, even at levels
significantly more sensitive than we can probe with the current data.

\begin{table*}
\caption{ Summary of full and partial branching fraction systematic
    errors (\%) for the $D^0 \to \pi^- e^+ \nu_e$ and $D^0 \to K^- e^+ \nu_e$ signal decay modes. The sign represents the direction of change relative to the change in the $[0,0.4)$ GeV$^2/c^4$ interval in each mode.}
\label{table:syst_d0}
\begin{ruledtabular}
\begin{tabular}{cdddddddddddd}
 & \multicolumn{6}{c}{$D^0 \to \pi^- e^+ \nu_e$} & \multicolumn{6}{c}{$D^0 \to K^- e^+ \nu_e$}\\
 & \multicolumn{6}{c}{$q^2$ interval (GeV$^2/c^4$)} &
\multicolumn{6}{c}{$q^2$ interval (GeV$^2/c^4$)} \\
Systematic  & \multicolumn{1}{c}{$[0,0.4)$} & \multicolumn{1}{c}{$[0.4,0.8)$} & \multicolumn{1}{c}{$[0.8,1.2)$} & \multicolumn{1}{c}{$[1.2,1.6)$}
            & \multicolumn{1}{c}{$[1.6,q^2_{\mathrm{max}}]$} & \multicolumn{1}{c}{All $q^2$} 
            & \multicolumn{1}{c}{$[0,0.4)$} & \multicolumn{1}{c}{$[0.4,0.8)$} & \multicolumn{1}{c}{$[0.8,1.2)$} & \multicolumn{1}{c}{$[1.2,1.6)$} 
            & \multicolumn{1}{c}{$[1.6,q^2_{\mathrm{max}}]$} & \multicolumn{1}{c}{All $q^2$} \\
Number $D\bar{D}$                      & 1.51 &  1.51 &  1.51 &  1.51 &  1.51 &  1.51 &  1.51 &  1.51 &  1.51 &  1.51 &  1.51 &  1.51 \\
$\nu_e$ simulation                     & 1.45 &  1.77 &  2.21 &  2.87 &  1.59 &  1.69 &  1.52 &  1.99 &  1.96 &  2.39 &  1.28 &  1.80 \\
$\pi^0$ efficiency                     & 0.01 &  0.01 &  0.01 & 0.00  & -0.03 & -0.00 &  0.00 &  0.00 &  0.00 &  0.00 &  0.01 &  0.00 \\
$K_S^0$ efficiency                     & 0.01 & -0.01 & -0.02 & -0.04 & -0.08 & -0.03 &  0.01 &  0.00 &  0.00 &  0.00 &  0.07 &  0.01 \\
$\pi^-$ PID                            & 0.44 &  0.41 &  0.39 &  0.39 &  0.40 &  0.41 &  0.00 &  0.00 &  0.00 &  0.00 &  0.00 &  0.00 \\
$K^-$ PID                              & 0.05 &  0.02 &  0.02 &  0.01 &  0.02 &  0.03 &  0.32 &  0.30 &  0.28 &  0.26 &  0.26 &  0.30 \\
$e^+$ PID                              & 0.76 &  0.40 &  0.75 &  0.36 &  0.44 &  0.56 &  0.65 &  0.58 &  0.60 &  0.52 &  0.46 &  0.61 \\
$e^+$ Fakes                            & 2.50 &  0.45 & -0.01 &  0.05 &  0.92 &  0.88 &  0.57 &  0.07 & -0.04 &  0.01 &  0.39 &  0.25 \\
$\pi^0$ production                     & 0.01 &  0.02 &  0.03 &  0.01 &  0.06 &  0.03 &  0.00 &  0.00 & -0.01 &  0.00 &  0.01 &  0.00 \\
$\pi^-$ production                     & 0.07 &  0.42 &  0.44 &  0.10 & -1.96 & -0.24 &  0.00 &  0.00 &  0.00 &  0.05 &  0.23 &  0.01 \\
$K^-$ fakes                            & 0.67 &  1.01 &  0.93 &  0.35 & -0.08 &  0.58 &  0.00 &  0.00 & -0.01 & -0.01 & -0.02 &  0.00 \\
$\pi^-e^+\nu_e$ $M_{\mathrm{bc}}$ res. & 0.93 &  1.06 &  0.85 &  0.83 &  1.05 &  0.95 &  0.02 &  0.01 &  0.01 &  0.01 &  0.06 &  0.01 \\
$K^-e^+\nu_e$ $M_{\mathrm{bc}}$ res.   & 0.07 &  0.04 &  0.02 &  0.01 &  0.01 &  0.03 &  0.09 &  0.08 &  0.07 &  0.09 &  0.16 &  0.09 \\
$\pi^0e^+\nu_e$ $M_{\mathrm{bc}}$ res. & 0.01 &  0.02 &  0.00 &  0.01 & -0.13 & -0.02 &  0.00 &  0.00 &  0.00 &  0.00 &  0.00 &  0.00 \\
$e^+$ veto                             & 0.05 &  0.04 & -0.01 & -0.14 & -0.01 &  0.00 &  0.08 &  0.07 &  0.00 & -0.02 & -0.03 &  0.05 \\
FSR                                    & 0.85 &  1.53 &  0.97 &  0.91 &  0.75 &  0.99 &  0.63 &  0.61 &  0.56 &  0.48 &  0.47 &  0.59 \\
Model dep.                             & 0.50 & -0.01 & -0.09 &  0.43 & -1.55 & -0.19 &  0.33 & -0.11 & -0.16 & -0.41 & -1.29 &  0.02 \\
\bfseries Total                        & 3.70 & 3.26  &  3.26 &  3.56 &  3.74 &  2.95 &  2.44 &  2.66 &  2.63 &  2.95 &  2.51 &  2.53 \normalfont \\
\end{tabular}
\end{ruledtabular}
\end{table*}

\begin{table*}
\caption{ Summary of full and partial branching fraction systematic
    errors (\%) associated with neutrino modeling in the MC 
    for the $D^0 \to \pi^- e^+ \nu_e$ and $D^0 \to K^- e^+ \nu_e$ signal decay modes. The sign represents the direction of change relative to the change in the $[0,0.4)$ GeV$^2/c^4$ interval in each mode.}
\label{table:nu_syst_d0}
\begin{ruledtabular}
\begin{tabular}{cdddddddddddd}
 & \multicolumn{6}{c}{$D^0 \to \pi^- e^+ \nu_e$} & \multicolumn{6}{c}{$D^0 \to K^- e^+ \nu_e$}\\
 & \multicolumn{6}{c}{$q^2$ interval (GeV$^2/c^4$)} & \multicolumn{6}{c}{$q^2$ interval (GeV$^2/c^4$)}\\
$\nu$ Systematic  & \multicolumn{1}{c}{$[0,0.4)$} & \multicolumn{1}{c}{$[0.4,0.8)$} & \multicolumn{1}{c}{$[0.8,1.2)$} & \multicolumn{1}{c}{$[1.2,1.6)$}
            & \multicolumn{1}{c}{$[1.6,q^2_{\mathrm{max}}]$} & \multicolumn{1}{c}{All $q^2$} 
            & \multicolumn{1}{c}{$[0,0.4)$} & \multicolumn{1}{c}{$[0.4,0.8)$} & \multicolumn{1}{c}{$[0.8,1.2)$} & \multicolumn{1}{c}{$[1.2,1.6)$} 
            & \multicolumn{1}{c}{$[1.6,q^2_{\mathrm{max}}]$} & \multicolumn{1}{c}{All $q^2$} \\
split-off showers    & 0.58 & 0.90 & 1.92 & 2.58 & 0.48 & 1.17 & 0.89 & 1.54 & 1.49 & 2.00 & 0.67 & 1.29 \\
$K_L^0$ showers      & 0.15 & 0.13 & 0.11 & 0.06 &-0.60 &-0.04 & 0.01 &-0.02 &-0.04 &-0.05 &-0.06 &-0.01 \\
$K_L^0$ production   & 0.67 & 0.66 & 0.66 & 0.67 & 0.65 & 0.66 & 0.66 & 0.65 & 0.66 & 0.67 & 0.69 & 0.66 \\
track efficiency     & 0.60 & 0.56 & 0.50 & 0.63 & 0.45 & 0.54 & 0.37 & 0.39 & 0.44 & 0.44 & 0.21 & 0.39 \\
track resolution     & 0.00 & 1.02 & 0.01 & 0.46 & 0.85 & 0.46 & 0.28 & 0.43 & 0.50 & 0.60 & 0.34 & 0.39 \\
split-off rejection  & 0.58 &-0.02 & 0.16 & 0.04 &-0.22 & 0.12 & 0.59 & 0.56 & 0.48 & 0.43 & 0.21 & 0.54 \\
particle ID          & 0.01 &-0.02 & 0.08 & 0.02 & 0.09 & 0.04 & 0.06 & 0.04 & 0.05 & 0.03 & 0.04 & 0.05 \\
shower resolution    & 0.03 & 0.09 & 0.06 &-0.01 & 0.11 & 0.06 & 0.00 & 0.00 & 0.00 & 0.00 &-0.01 & 0.00 \\
fake tracks          & 0.76 & 0.71 & 0.70 & 0.71 & 0.72 & 0.72 & 0.72 & 0.72 & 0.71 & 0.71 & 0.71 & 0.72 \\
\bfseries Total      & 1.45 & 1.77 & 2.21 & 2.87 & 1.59 & 1.69 & 1.52 & 1.99 & 1.96 & 2.39 & 1.28 & 1.80 \normalfont \\
\end{tabular}
\end{ruledtabular}
\end{table*}

\begin{table*}
\caption{ Summary of full  and partial branching fraction systematic
    errors (\%) for the $D^+ \to \pi^0 e^+ \nu_e$ and $D^+ \to \bar{K}^0 e^+\nu_e$ signal decay modes. The sign represents the direction of change relative to the change in the $[0,0.4)$ GeV$^2/c^4$ interval in each mode. }
\label{table:syst_dp}
\begin{ruledtabular}
\begin{tabular}{cdddddddddddd}
 & \multicolumn{6}{c}{$D^+ \to \pi^0 e^+ \nu_e$} & \multicolumn{6}{c}{$D^+ \to \bar{K}^0 e^+ \nu_e$}\\
 & \multicolumn{6}{c}{$q^2$ interval (GeV$^2/c^4$)} & \multicolumn{6}{c}{$q^2$ interval (GeV$^2/c^4$)}\\
Systematic  & \multicolumn{1}{c}{$[0,0.4)$} & \multicolumn{1}{c}{$[0.4,0.8)$} & \multicolumn{1}{c}{$[0.8,1.2)$} & \multicolumn{1}{c}{$[1.2,1.6)$}
            & \multicolumn{1}{c}{$[1.6,q^2_{\mathrm{max}}]$} & \multicolumn{1}{c}{All $q^2$} 
            & \multicolumn{1}{c}{$[0,0.4)$} & \multicolumn{1}{c}{$[0.4,0.8)$} & \multicolumn{1}{c}{$[0.8,1.2)$} & \multicolumn{1}{c}{$[1.2,1.6)$} 
            & \multicolumn{1}{c}{$[1.6,q^2_{\mathrm{max}}]$} & \multicolumn{1}{c}{All $q^2$} \\
Number $D\bar{D}$                      & 1.60 & 1.60 & 1.60 & 1.60 & 1.60 & 1.60 & 1.60 & 1.60 & 1.60 & 1.60 & 1.60 & 1.60 \\
$\nu$ simulation                       & 2.54 & 3.41 & 2.57 & 2.53 & 2.45 & 1.96 & 1.71 & 1.75 & 1.82 & 1.84 & 2.18 & 1.74 \\
$\pi^0$ efficiency                     & 0.87 & 0.56 & 0.77 & 1.07 & 1.07 & 0.85 & 0.00 & 0.01 &-0.01 &-0.02 & 0.00 & 0.00 \\
$K_S^0$ efficiency                     & 0.02 & 0.02 & 0.08 & 0.11 & 0.18 & 0.07 & 1.05 & 1.00 & 0.94 & 0.88 & 0.81 & 1.00 \\
$\pi^-$ PID                            & 0.17 & 0.37 & 0.13 &-0.24 &-0.29 & 0.06 & 0.01 & 0.01 & 0.00 &-0.03 &-0.02 & 0.00 \\
$K^-$ PID                              & 0.17 & 0.37 & 0.12 &-0.24 &-0.29 & 0.06 & 0.01 & 0.01 & 0.00 &-0.02 & 0.00 & 0.00 \\
$e^+$ PID                              & 1.13 & 0.56 & 0.33 & 0.98 & 0.01 & 0.62 & 0.62 & 0.65 & 0.52 & 0.59 & 0.76 & 0.61 \\
$e^+$ Fakes                            & 1.52 & 0.14 &-0.29 &-0.07 & 0.64 & 0.44 & 0.38 &-0.03 &-0.17 & 0.09 & 1.00 & 0.14 \\
$\pi^0$ production                     & 0.43 & 0.81 & 0.76 &-0.73 &-1.87 &-0.04 & 0.02 &-0.01 & 0.00 &-0.11 &-0.14 &-0.01 \\
$\pi^-$ production                     & 0.07 &-0.02 & 0.03 & 0.02 & 1.46 & 0.29 & 0.02 &-0.01 & 0.00 & 0.19 & 1.12 & 0.04 \\
$K^-$ fakes                            & 0.01 &-0.01 & 0.04 & 0.07 & 0.20 & 0.05 & 0.01 & 0.01 &-0.03 &-0.07 &-0.16 &-0.01 \\
$\pi^-e^+\nu_e$ $M_{\mathrm{bc}}$ res. & 0.00 & 0.00 &-0.01 & 0.05 &-0.29 &-0.05 & 0.01 & 0.01 & 0.00 & 0.04 & 0.09 & 0.01 \\
$K^-e^+\nu_e$ $M_{\mathrm{bc}}$ res.   & 0.00 & 0.01 & 0.00 & 0.00 & 0.01 & 0.00 & 0.02 & 0.01 & 0.00 & 0.00 & 0.02 & 0.01 \\
$\pi^0e^+\nu_e$ $M_{\mathrm{bc}}$ res. & 2.62 & 1.27 & 3.77 & 1.17 & 1.97 & 2.09 & 0.01 & 0.01 & 0.01 & 0.05 & 0.08 & 0.01 \\
$e^+$ veto                             & 0.26 &-0.01 & 0.20 & 0.03 &-0.14 & 0.07 & 0.02 & 0.09 & 0.02 & 0.12 &-0.30 & 0.04 \\
FSR                                    & 0.26 & 0.48 & 0.47 & 0.68 & 0.65 & 0.49 & 0.25 & 0.46 & 0.55 & 0.64 & 0.60 & 0.41 \\
Model dep.                             & 0.56 & 0.08 &-0.08 & 0.76 & 0.08 & 0.28 & 0.35 &-0.16 &-0.28 &-0.83 &-1.51 &-0.06 \\
\bfseries Total                        & 4.57 & 4.19 & 5.00 & 3.76 & 4.52 & 3.53 & 2.70 & 2.70 & 2.73 & 2.87 & 3.69 & 2.67 \normalfont \\
\end{tabular}
\end{ruledtabular}
\end{table*}

\begin{table*}
\caption{ Summary of full and partial branching fraction systematic
    errors (\%) associated with neutrino modeling in the MC
    for the $D^+ \to \pi^0 e^+ \nu_e$ and $D^+ \to \bar{K}^0 e^+\nu_e$ signal decay modes. The sign represents the direction of change relative to the change in the $[0,0.4)$ GeV$^2/c^4$ interval in each mode.}
\label{table:nu_syst_dp}
\begin{ruledtabular}
\begin{tabular}{cdddddddddddd}
 & \multicolumn{6}{c}{$D^+ \to \pi^0 e^+ \nu_e$} & \multicolumn{6}{c}{$D^+ \to \bar{K}^0 e^+ \nu_e$}\\
 & \multicolumn{6}{c}{$q^2$ interval (GeV$^2/c^4$)} & \multicolumn{6}{c}{$q^2$ interval (GeV$^2/c^4$)}\\
$\nu$ Systematic  & \multicolumn{1}{c}{$[0,0.4)$} & \multicolumn{1}{c}{$[0.4,0.8)$} & \multicolumn{1}{c}{$[0.8,1.2)$} & \multicolumn{1}{c}{$[1.2,1.6)$}
            & \multicolumn{1}{c}{$[1.6,q^2_{\mathrm{max}}]$} & \multicolumn{1}{c}{All $q^2$} 
            & \multicolumn{1}{c}{$[0,0.4)$} & \multicolumn{1}{c}{$[0.4,0.8)$} & \multicolumn{1}{c}{$[0.8,1.2)$} & \multicolumn{1}{c}{$[1.2,1.6)$} 
            & \multicolumn{1}{c}{$[1.6,q^2_{\mathrm{max}}]$} & \multicolumn{1}{c}{All $q^2$} \\
split-off showers      & 0.62 & 2.94 & 2.11 & 1.30 &-1.68 & 1.14 & 0.17 & 0.37 & 0.21 &-0.13 &-1.36 & 0.18 \\
$K_L^0$ showers        & 0.19 & 0.17 & 0.08 & 0.13 &-0.83 &-0.03 & 0.01 & 0.06 &-0.08 &-0.15 &-0.46 &-0.02 \\
$K_L^0$ production     & 1.10 & 1.07 & 1.07 & 1.08 & 1.13 & 1.09 & 1.07 & 1.07 & 1.09 & 1.10 & 1.09 & 1.08 \\
track efficiency       & 0.51 & 0.37 & 0.18 &-0.14 & 0.13 & 0.24 & 0.62 & 0.57 & 0.70 & 0.66 & 0.39 & 0.62 \\
track resolution       & 1.08 & 0.05 &-0.09 &-0.90 & 0.13 & 0.12 & 0.43 & 0.45 & 0.64 & 0.49 & 0.92 & 0.49 \\
split-off rejection    & 1.66 & 1.08 & 0.68 & 1.45 &-0.81 & 0.86 & 0.81 & 0.88 & 0.84 & 0.99 & 0.21 & 0.84 \\
particle ID            & 0.09 &-0.02 & 0.04 & 0.07 &-0.03 & 0.03 & 0.01 & 0.02 & 0.00 & 0.01 & 0.08 & 0.01 \\
shower resolution      & 0.00 &-0.01 & 0.18 & 0.30 & 0.04 & 0.08 & 0.00 &-0.02 & 0.03 &-0.01 &-0.09 & 0.00 \\
fake tracks            & 0.77 & 0.71 & 0.70 & 0.71 & 0.71 & 0.72 & 0.72 & 0.70 & 0.69 & 0.69 & 0.70 & 0.71 \\
\bfseries Total        & 2.54 & 3.41 & 2.57 & 2.53 & 2.45 & 1.96 & 1.71 & 1.75 & 1.82 & 1.84 & 2.18 & 1.74 \normalfont \\
\end{tabular}
\end{ruledtabular}
\end{table*}

\begin{table*}[t]
\caption{Summary of the efficiencies ($\varepsilon$) and efficiency-corrected
  yields for each $q^2$ interval and the corresponding partial branching fractions, 
  the total branching fractions, the branching ratios and the isospin ratios. In all cases the
  first errors are statistical and the second are systematic. For the $\bar{K}^0$ mode, the efficiency and
  yields correspond to the reconstructed $K^0_S\to\pi^+\pi^-$ mode, so do not include the initial production amplitude or $\pi^+\pi^-$ branching fraction factors.}
\label{table:bf}
\begin{ruledtabular}
\begin{tabular}{lllllll}
\multicolumn{7}{c}{$q^2$ interval (GeV$^2/c^4$)} \\
                 & $< 0.4$                 & $0.4
- 0.8$           & $0.8 - 1.2$          & $1.2 - 1.6$ 
& $ \ge 1.6 $          & Total \\
\hline
\multicolumn{7}{c}{$D^0 \to \pi^- e^+ \nu_e$} \\
$\varepsilon$ (\%) & $19.4$ & $21.0$ & $22.4$ & $22.8$ & $22.4$ & --  \\
Yield & $1452(113)(49)$ & $1208(102)(35)$ & $1242(99)(36)$ & $906(85)(29)$ & $1357(103)(46)$
& -- \\
${\cal B}(\pi^- e^+ \nu_e)(\%)$         & $0.070(5)(3)$      & $0.059(5)(2)$
& $0.060(5)(2)$    & $0.044(4)(2)$       & $0.066(5)(2)$      &
$0.299(11)(9)$ \\
\hline
\multicolumn{7}{c}{$D^+ \to \pi^0 e^+ \nu_e$} \\
$\varepsilon$ (\%) & $7.5$ & $8.0$ & $7.9$  & $7.2$ & $5.7$ & --  \\
Yield & $1379(168)(59)$ & $1584(180)(61)$ & $1012(154)(48)$  & $1028(158)(35)$ & $1101(174)(47)$
& -- \\
${\cal B}(\pi^0 e^+ \nu_e)(\%)$         & $0.084(10)(4)$ & $0.097(11)(4)$ &
$0.062(9)(3)$ & $0.063(10)(2)$ & $0.067(11)(3)$ & $0.373(22)(13)$ \\
\hline
\multicolumn{7}{c}{$D^0 \to K^- e^+ \nu_e$} \\
$\varepsilon$ (\%) & $19.2$ & $20.5$ & $20.0$ & $18.3$ & $13.9$ & -- \\
Yield & $29701(441)(569)$ & $21600(377)(473)$ & $14032(304)(301)$ & $7001(225)(178)$ & $991(112)(20)$
& -- \\
${\cal B}(K^- e^+ \nu_e)(\%)$           & $1.441(21)(35)$        & $1.048(18)(28)$ &
$0.681(15)(18)$ & $0.340(11)(10)$  & $0.048(5)(12)$   & $3.557(33)(90)$ \\
\hline
\multicolumn{7}{c}{$D^+ \to \bar{K}^0 e^+ \nu_e$} \\
$\varepsilon$ (\%) & $11.7$ & $12.3$ & $12.5$ & $12.2$ & $12.5$ & -- \\
Yield & $19480(466)(417)$ & $14422(415)(306)$ & $9009(327)(194)$ & $4656(236)(107)$ & $789(104)(26)$
& -- \\
${\cal B}(\bar{K}^0 e^+ \nu_e)(\%)$ & $3.436(82)(93)$   & $2.544(73)(69)$         &
$1.589(58)(44)$ & $0.821(42)(24)$ & $0.139(18)(5)$ & $8.53(13)(23)$ \\
\hline
& & & & & & \\
$R_0 (\%)$                                & $4.89(39)(12)$   &
$5.59(48)(12)$    & $8.85(74)(15)$ & $12.9(13)(2)$    & $137(19)(3)$ &
$8.41(32)(13)$ \\
$R_+ (\%)$                                & $2.45(31)(9)$   &
$3.80(45)(13)$    & $3.89(61)(17)$ & $7.6(12)(2)$      & $48(10)(2)$ &
$4.37(27)(12)$ \\
$I_\pi$         & $2.12(31)(9)$    & $1.54(22)(7)$    & $2.47(43)(13)$ & $1.78(32)(7)$     & $2.48(45)(13)$ & $2.03(14)(8)$ \\
$I_K$           & $1.06(3)(3)$      & $1.04(4)(3)$     & $1.09(5)(3)$   & $1.05(6)(4)$      & $0.88(15)(3)$  & $1.06(2)(3)$ \\
\end{tabular}
\end{ruledtabular}
\end{table*}

\section{Experimental Systematic Uncertainties}
\label{sec:syst_errors}

The systematic uncertainties for the $D^0$ modes are summarized in
Tables~\ref{table:syst_d0} and~\ref{table:nu_syst_d0}. The first table
presents the complete list, while the second breaks down the 
neutrino-reconstruction simulation errors into component parts.  The corresponding
systematics tables for the $D^+$ modes are presented in 
Tables~\ref{table:syst_dp} and~\ref{table:nu_syst_dp}.
For individual uncertainties we give the sign of the error relative to the change in the
lowest $q^2$ range. The largest systematic
uncertainties are those associated with the number of  $D\bar{D}$ pairs (needed for normalization in 
the branching fraction determination, as described in Section VI)
and with neutrino reconstruction simulation.    
Uncertainties in neutrino simulation include both inaccuracies in detector simulation and
uncertainty in the decay model of the non-signal $D$, as discussed
above. 

The starting point for the assessment of many of the systematic
uncertainties is the measurement of any discrepancies between data
and MC in the desired quantities (e.g., signal pion efficiency, signal
kaon efficiency, etc.). Such measurements (or limits) are made using an independent
data sample - in most cases, a sample of events with
one of the two $D$ mesons from the $\psi(3770)$ fully reconstructed in a
hadronic mode. In the case of significant discrepancies, the MC samples are
corrected for use in our nominal fit (the fit used to obtain our final
branching fraction results, as opposed to any of the fits used to obtain
systematic uncertainties) as noted above. Such corrections
lead to changes in the measured yields of up to a few
percent, but are determined precisely enough to yield sub-percent
systematic uncertainties. For each
systematic category, we determine the size of its contribution by
biasing the MC samples away from their
nominal configuration at the level given by the uncertainty of the
independent study. We re-fit the data with these biased MC samples, and
use the deviation of the fit results from their nominal values to provide
an estimate of the uncertainty.  We note that because of the correlations
among the five $q^2$ intervals in a given mode, the sum over $q^2$ of
the systematic errors tends to be less sensitive to the systematic
variations than the individual intervals themselves.

The number of $D\bar{D}$ pairs, used to convert the measured yields to
branching fractions (see below), is a direct product of the CLEO-c
hadronic branching fraction analysis~\cite{Dobbs:2007zt}.  We combine the
statistical and systematic uncertainties from that analysis for our uncertainty estimates.

We have assessed the uncertainties associated with the 
finding and identification efficiency for each of the signal hadrons. For the
signal $K^\pm$ and
$\pi^\pm$,  the charged track-finding efficiency is already accounted in the
tracking efficiency portion of the $\nu$ simulation uncertainty. They have, however, 
additional particle identification (PID) criteria associated with them, for which we assess a
correction and uncertainty. For the signal $\pi^0$ and $K_S^0$ we
assess a correction and uncertainty for the reconstruction efficiencies of these
particles. We evaluate each of these four uncertainties by first measuring
a momentum dependent, and hence $q^2$ dependent, correction and fit the
measurements with a linear parameterization.  The best fit result is
applied as a correction in the nominal fit.  To evaluate the
systematic uncertainty, we identify the largest systematic variation on
the $\chi^2=1$ ellipse from the linear fit.  The branching fractions are
most affected by the largest variation in overall normalization, while
the form factors (Section~\ref{sec:ff}) are most affected by the largest variation in slope.

We have uncertainties associated with the electron identification efficiency and
the rates for hadrons to mis-reconstruct as (fake) electrons.  We vary the efficiency and
fake rates used in the analysis of our MC samples (see Section~\ref{sec:fit_comp}) according to
 the uncertainties from the data studies used to measure them, and re-fit the data to
 evaluate our sensitivity.

Modeling of $\pi^\pm$ and $\pi^0$ production --
spectra and rates -- in $D$ decay significantly affects the background shape and
rate for the signal cross-feed background into the pion signal modes. The 
large effect results because a pion from the non-signal $D$ decay can be
swapped in as the signal pion candidate. We measure the background pion
spectra in data using the inclusive $D$ decays on the ``other side'' of a fully reconstructed
hadronically-decayed $D$ ``tag'', and correct the MC spectra accordingly. To be
conservative in the associated systematic
uncertainty,  we take the full difference for results obtained using corrected and uncorrected
spectra. In the signal $D^0 \to \pi^- e^+
\nu_e$ mode we also correct the cross-feed background from $D^0 \to K^-
e^+ \nu_e$ events that results from misidentifying a
$K^\pm$ as a $\pi^\pm$. Once again the
uncertainty estimate is taken as the difference of our measured rates
obtained using the corrected and
uncorrected fake rates.

For three of our signal modes, $D^0 \to \pi^- e^+ \nu_e$, $D^0 \to
K^- e^+ \nu_e$ and $D^+ \to \pi^0 e^+ \nu_e$, we have systematic uncertainty
associated with the additional $M_{\mathrm{bc}}$ resolution parameter.  The
statistical uncertainty already has a contribution from allowing this parameter to float.
We estimate the contribution to the systematic uncertainty for each mode
by increasing the value of that mode's resolution parameter by one standard deviation beyond the best fit result.

We must also account for any uncertainty associated with modeling
event loss from the single electron veto because of secondary electrons from
photon conversions and other processes. According to data studies (using
the CLEO-c ``tagged'' samples, where one of the two $D$ mesons from the
$\psi(3770)$ is fully reconstructed), 
our MC simulation models the number of
secondary electrons in our events accurately within the error of the
study. The most likely potential source of uncertainty arises from mis-modeling the rate
 for photon conversion within the detector material. For the uncertainty estimate we therefore 
 vary this contribution over the range allowed by the maximum allowed uncertainty of our data
study, about 8\%.  
 
For the systematic uncertainty associated with the final state radiation (FSR) modeling, we
take the difference between the KLOR and PHOTOS predictions. This
simulates a change in the radiative branching fraction of up to 16\% in
the most extreme case. Because the majority of the correction results from the lack of
the FSR interference terms between the charged hadron and electron, the systematic
should be an overestimate
of the FSR uncertainty from final or initial-state particles. This overestimate
compensates for the unknown direct (structure-dependent) contributions.  

The final systematic error we assess is the dependence on our modeling of the form factor input to our signal MC. We
reweight each of our signal MC samples with a different form factor
input, namely ISGW2. The nominal form factor input to our signal MC is a
BK parameterization~\cite{BKparam} with parameters determined by lattice
QCD~\cite{fnalqcd}.  The $q^2$ spectra of the latter differ markedly from
those of ISGW2. We fit with the re-weighted MC spectra and the
difference to the nominal fit gives the systematic error associated with  model
dependence.  The small uncertainties obtained in this study confirm our conclusion drawn from
fitting the large MC sample.
   
\section{Branching Fraction Results}
\label{sec:bf_results}
Combining the results of the fit and the systematic uncertainty estimates gives us the
final efficiency-corrected yield measurement for each mode. From that yield ($Y$), 
we obtain the branching
fraction ${\cal B} = Y\slash 2N_{D\bar{D}}$,
where $N_{D\bar{D}}$ is the number of neutral ($N_{D^0\bar{D}^0}$) or
charged ($N_{D^+D^-}$) pairs in our sample.   
We obtain these numbers from an independent CLEO-c analysis~\cite{Dobbs:2007zt}
based on the comparison of events with one reconstructed $D$ to events
with both $D$ decays reconstructed, in certain hadronic modes.
For the same data set that we have used, that analysis finds
$N_{D^0\bar{D}^0} = (1.031 \pm 0.016) \times 10^6$
and 
$N_{D^+D^-} = (0.819 \pm 0.013) \times 10^6$.
Our fit yields, efficiencies and branching
fractions for each mode, in each $q^2$ range, are presented in
Table~\ref{table:bf}. The total 
branching fractions for each mode (also listed in Table~\ref{table:bf}) are
\begin{equation}
{\cal B}(D^0 \to \pi^- e^+ \nu_e) = 0.299(11)(9)\%,
\end{equation}
\begin{equation}
{\cal B}(D^+ \to \pi^0 e^+ \nu_e) = 0.373(22)(13)\%,
\end{equation}
\begin{equation}
{\cal B}(D^0 \to K^- e^+ \nu_e) = 3.56(3)(9)\%,
\end{equation}
and
\begin{equation}
{\cal B}(D^+ \to \bar{K}^0 e^+ \nu_e) = 8.53(13)(23)\%.
\end{equation}
The errors listed are statistical and systematic, respectively.

We also measure the branching fraction and partial width ratios in each
$q^2$ range. The full results are given in Table~\ref{table:bf}. 
To determine the partial width ratios we used the Particle Data Group
 lifetimes~\cite{pdg}
    $\tau_{D^0} = 410.3 \pm 1.5$ fs and $\tau_{D^+} =
    1040 \pm 7$ fs.
For the integrated $q^2$ ranges we find the ratios of branching fractions
\begin{equation}
R_0 \equiv \frac{{\cal B}(D^0 \to \pi^- e^+ \nu_e)}{{\cal B}(D^0 \to K^-
  e^+ \nu_e)} = 8.41(32)(13)\%
\end{equation}
and
\begin{equation}
R_+ \equiv \frac{{\cal B}(D^+ \to \pi^0 e^+ \nu_e)}{{\cal B}(D^+ \to
  \bar{K}^0 e^+ \nu_e)} = 4.37(27)(12)\%.
\end{equation}
The partial width ratios, which are expected to satisfy isospin relationships, are
found to be
\begin{equation}
I_\pi \equiv \frac{\Gamma(D^0 \to \pi^- e^+ \nu_e)}{\Gamma(D^+ \to \pi^0
  e^+ \nu_e)} = 2.03(14)(8)
\end{equation}
and
\begin{equation}
I_K \equiv \frac{\Gamma(D^0 \to K^- e^+ \nu_e)}{\Gamma(D^+ \to \bar{K}^0
  e^+ \nu_e)} = 1.06(2)(3).
\end{equation}
We expect $I_\pi = 2$ and $I_K = 1$, hence the measured partial width ratios satisfy
isospin symmetry within our experimental precision.

\section{Form factors}
\label{sec:ff}

\begin{table*}[t]
\caption{Summary of form factor results for the series parameterization
  and pole model fits. Correlation coefficients for the total uncertainty
  between variables in any
  two (three) preceding columns are given by $\rho$ ($\rho_{ij}$). The
  first errors are statistical and the second are systematic. The values for
  the $\pi^0 e^+ \nu_e$ mode are isospin corrected. For the
  series parameters ($a_i$) we have assumed $|V_{cs}| = 0.976$ and
  $|V_{cd}| = 0.224$. }
\label{table:all_ff}
\begin{ruledtabular}
\begin{tabular}{lllllllllll}
& \multicolumn{10}{c}{Series Parameterization - Three Parameter Fits} \\
Decay                 & $a_0$       & $a_1$ & $a_2$  & $\rho_{01}$ & $\rho_{02}$ & $\rho_{12}$ & $|V_{cq}|f_+(0)$ & $1 + 1\slash \beta - \delta$  & $\rho$ & $\chi^2/d.o.f$\\
$\pi^- e^+ \nu_e$     & 0.044(2)(1) & -0.18(7)(2)  & -0.03(35)(12)& 0.81$\;$ & 0.71$\;$ & 0.96$\;$ & 0.140(7)(3)  & 1.30(37)(12) & -0.85 & 2.0/2\\
$\pi^0 e^+ \nu_e$     & 0.044(3)(1) & -0.23(11)(2) & -0.60(57)(15)& 0.80     & 0.67 & 0.95         & 0.138(11)(4) & 1.58(60)(13) & -0.86 & 2.8/2 \\
$K^- e^+ \nu_e$       & 0.0234(3)(3)& -0.009(21)(7)&  0.52(28)(6) & 0.62     & 0.56 & 0.96         & 0.747(9)(9)  & 0.62(13)(4)  & -0.62 & 0.2/2 \\
$\bar{K}^0 e^+ \nu_e$ & 0.0224(4)(3)& 0.009(32)(7) &  0.76(42)(8) & 0.72     & 0.64 & 0.96         & 0.733(14)(11)& 0.51(20)(4)  & -0.72 & 1.7/2\\
\hline
& \multicolumn{10}{c}{Series Parameterization - Two Parameter Fits} \\
Decay & $a_0$ & $a_1$ & $\rho$ & \multicolumn{3}{l}{$|V_{cq}|f_+(0)$} & $1+1\slash\beta - \delta$ & $\rho$ & $\chi^2/d.o.f$\\
$\pi^- e^+ \nu_e$ & $0.044(2)(1)$ & -$0.173(19)(7)$  & 0.66 & \multicolumn{3}{l}{$0.140(5)(3)$} & $1.27(11)(4)$ & -0.80 & 2.0/3 \\
$\pi^0 e^+ \nu_e$ & $0.046(2)(1)$ & -$0.124(30)(9)$ & 0.69  & \multicolumn{3}{l}{$0.147(7)(4)$} & $1.01(16)(5)$  & -0.78 & 4.0/3 \\
$K^- e^+ \nu_e$ & $0.0230(2)(3) $ & -$0.047(6)(3)$ & 0.34 & \multicolumn{3}{l}{$0.734(6)(9)$} & $0.86(4)(2)$ & -0.43 & 3.8/3 \\
$\bar{K}^0 e^+ \nu_e$ & $0.0218(3)(3)$ & -$0.046(9)(4)$ & 0.53 &\multicolumn{3}{l}{$0.713(9)(11)$} & $0.87(6)(3)$ & -0.60 & 4.9/3 \\
\hline
& \multicolumn{6}{c}{Simple Pole Model Fits} & \multicolumn{4}{c}{Modified Pole Model Fits} \\
Decay & $|V_{cq}|f_+(0)$ & $m_{\mathrm{pole}}$ (GeV$/c^2$) & $\rho$ & \multicolumn{3}{l}{$\chi^2/d.o.f$} & $|V_{cq}|f_+(0)$ & $\alpha$ & $\rho$ & $\chi^2/$ d.o.f\\
$\pi^- e^+ \nu_e$     & 0.146(4)(2) & 1.87(3)(1) & 0.63 & \multicolumn{3}{l}{3.11/3} & 0.142(4)(2) & 0.37(8)(3) & -0.75 & 2.1/3\\
$\pi^0 e^+ \nu_e$     & 0.149(6)(3) & 1.97(7)(2) &  0.65 & \multicolumn{3}{l}{4.42/3} & 0.147(7)(4) & 0.14(16)(4) & -0.75 & 4.07/3\\
$K^- e^+ \nu_e$       & 0.735(5)(9) & 1.97(3)(1) & 0.36 & \multicolumn{3}{l}{2.67/3} & 0.732(6)(9) & 0.21(5)(3) & -0.42 & 4.32/3\\
$\bar{K}^0 e^+ \nu_e$ & 0.710(8)(10) & 1.96(4)(2) & 0.53 & \multicolumn{3}{l}{4.1/3} & 0.708(9)(10) & 0.22(8)(3) & -0.59 & 5.3/3\\
\end{tabular}
\end{ruledtabular}
\end{table*}

For each of our four signal decay modes we have obtained partial
branching fraction results in five $q^2$ ranges. To extract information
about the form factors we use the relationship 
\begin{equation}
\label{bfi}
{\cal B}_i =
\frac{G_F^2|V_{cq}|^2}{24\pi^3\Gamma_{D}}
\int_{q^2_{\mathrm{min}}(i)}^{q^2_{\mathrm{max}}(i)} 
p^3|f_{+}(q^2)|^2 dq^2
\end{equation}
to relate
the form factor $f_+(q^2)$ to the partial branching
fraction ${\cal B}$ in a particular $q^2$ range.
In this expression, $\Gamma_{D}$ is the total decay width of the parent
$D$ meson, and $i$ denotes the particular $q^2$ interval.
 A specific functional form is chosen for $f_+(q^2)$ 
(see Section~\ref{sec:semilep}) and the parameter values are determined 
via a $\chi^2$ fit to the five measured ${\cal B}_i$. In order to account for the 
correlations between the branching fractions in each $q^2$ range we
minimize the expression
\begin{equation}
\chi^2 = \sum_{ij} \left( {\cal B}_i - y_i \right)C_{ij}^{-1}\left( {\cal B}_j - y_j \right),
\end{equation}
where $y_i$ is the fit prediction for the branching fraction in the $i^{\mathrm{th}}$ $q^2$ 
interval,
and $C_{ij}^{-1}$ is the inverse of the covariance matrix. The
integration in each bin is performed numerically on each fit iteration using the
trapezoidal rule. 

The systematic uncertainties on the form factor parameters are evaluated using
the same method as for the branching fraction
analysis. We take the set of branching fractions that result from the branching
fraction fit for each systematic uncertainty, then redo the fit for the form factors. 
The difference in these fit parameters from the nominal results is taken as the
estimate of the systematic uncertainty.  The list of systematic uncertainties
evaluated is the same as for  the branching fraction analysis 
(see Section~\ref{sec:syst_errors}). 
Note that for the systematic errors found by exploring a
one standard deviation $\chi^2$ ellipse in normalization versus $q^2$-dependence and
taking the largest observed deviation from
the nominal result ($\pi^-$ PID, $K^-$ PID, $\pi^0$ finding, and $K^0_S$
finding), the form factors and the branching fractions will have their
largest deviations in very different regions of the ellipse. The form
factor will be most sensitive to the region of the ellipse
that causes the largest variation as a function of $q^2$, while the branching fraction
is most sensitive to the overall normalization.  

Fitting with the full covariance matrix that includes both statistical and systematic uncertainties and correlations (see Appendix~\ref{correl}) yields almost identical central values and total errors.

\begin{figure}[t]
\begin{center}
\hspace{0.0cm}
\includegraphics[width=8.5cm]{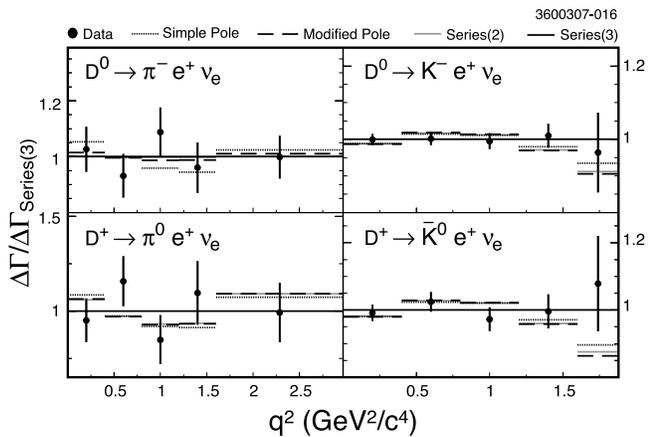} 
\caption{ Form factor fit comparison for all modes. All data (points) and fits (histograms) are
normalized to the relevant three-parameter series fit result (Series(3), line at 1).  
The simple pole, modified pole, and two-parameter series fit (Series(2)) are shown by
triple-dot-dash, dashed, and solid histograms, respectively.}
\label{fig:ff_all}
\end{center}
\end{figure}

We evaluate the form factor shape using the functional form given by the series
parameterization as described in Section~\ref{sec:semilep}. 
For comparative purposes we also provide results based on the
two pole models described in Section~\ref{sec:semilep}. 
For the series model we perform fits using both the first two and the first three
expansion parameters $a_k$. This tests both our sensitivity to the number of
parameters in the expansion and the convergence of the series. We express
our results in terms of the physical observables, the intercept $|V_{cq}|f_+(0)$ and
$1+1\slash\beta - \delta$, as well as giving the expansion parameters. 
In the simple pole model we fit for the
intercept and the pole mass $m_{\mathrm{pole}}$, while in the modified
pole model we fit for the intercept and the shape
parameter $\alpha$, which summarizes the effective pole
contribution. The results for all modes are summarized in Table~\ref{table:all_ff}. 
Comparisons of the four fits, for each of the four modes, are shown in
Fig.~\ref{fig:ff_all}. To allow systematic differences to be viewed
clearly between the various parameterizations, we normalize the data and
all fit results to the result of the three parameter series fit in each $q^2$ interval.

For the series expansion, comparison of the two-parameter and three-parameter fits
shows that our kaon data prefer a non-zero quadratic $z$
term. The probability of $\chi^2$ improves from 29\% (22\%) to 89\%
(44\%) going from two to three terms in the series for the $K^-$ ($K^0$)
fit. The pion measurements currently lack sensitivity to probe this term,
and two and three parameter fits yield similar results for the first two
parameters.  Since a quadratic term appears to be preferred for the kaons,
however, we include that term in our series fits to the pion data to
improve the probability that our shape uncertainties bracket the true
form factor shape.  While the central value for $a_2$ is an order of
magnitude larger than the other terms, we stress that regions of
parameter space with $a_2$ of similar magnitude to $a_0$ and $a_1$ fall
will within the 90\% hypercontour for the fit, so no strong statements
can be made about the size of $a_2$ or about the convergence (or
potential lack thereof) of the series from these data.

For the pole models we observe that the parameterizations can provide
a shape that describes our data adequately, but only with parameter values that do not support their
physical basis. Although the fits give quite reasonable $\chi^2$
values (see Table~\ref{table:all_ff}), the poles masses do not agree with the $M_{D_s^*}$
($M_{D^*}$) masses expected for the kaon (pion) modes by over $3\sigma$ 
for the most precise fits.  The $1+1\slash\beta -
\delta$ results from the $D^0 \to K^- e^+ \nu_e$ series expansion fit are over $3\sigma$ 
from the value of $\sim$2 necessary for physical
validity of the BK parameterization, while those derived from our
$\alpha$ values for the kaon modes are tens of $\sigma$ away.

\section{Extraction of $|V_{cs}|$ and $|V_{cd}|$}
\label{sec:ckm}

We extract $|V_{cd}|$ and $|V_{cs}|$ by combining our $|V_{cq}|f_+(0)$
results from the three parameter series expansion fits with the unquenched LQCD 
results \cite{fnalqcd} $f_+^{(D\to\pi)}(0)=0.64(3)(6)$ and 
 $f_+^{(D\to K)}(0)=0.73(3)(7)$. For the $D^0\to\pi^-$ and $D^+\to\pi^0$  modes we find 
$|V_{cd}| = 0.218 \pm 0.011 \pm 0.005 \pm 0.023$ and
$|V_{cd}| = 0.216 \pm 0.017 \pm 0.006 \pm 0.023$, respectively.
For the $D^0\to K^-$ and $D^+\to\bar{K}^0$ modes, we find
$|V_{cs}| = 1.023 \pm 0.013 \pm 0.013 \pm 0.107$ and
$|V_{cs}| = 1.004 \pm 0.020 \pm 0.015 \pm 0.105$.  
Averaging the $D^0$ and $D^+$ results (taking into account correlated and 
uncorrelated uncertainties) we find 
\begin{equation}
|V_{cd}| = 0.217 \pm 0.009 \pm 0.004 \pm 0.023
\end{equation}
 and 
\begin{equation}
|V_{cs}| = 1.015 \pm 0.010 \pm 0.011 \pm 0.106.
\end{equation}
The uncertainties, statistical, systematic and theoretical, respectively, 
are dominated by the discretization uncertainty in the LQCD charm quark action, 
which should be improved in the near future for the Fermilab action, or greatly reduced through the use of other actions..

We can also extract the ratio $|V_{cd}|/|V_{cs}|$ from the ratio of our
measured form factors.  From
the $z$ expansion fits to our $D^0$ data, we obtain 
$|V_{cd}|f_+^{(D\to\pi)}(0)/|V_{cs}|f_+^{(D\to K)}(0) = 0.187 \pm 0.010 \pm 0.003$,
while from our $D^\pm$ data we obtain
$|V_{cd}|f_+^{(D\to\pi)}(0)/|V_{cs}|f_+^{(D\to K)}(0) = 0.188 \pm 0.015 \pm 0.004$.  
The errors are statistical and systematic, respectively, and all
correlations have been taken into
account.  Averaging, again with correlated uncertainties accounted for, we obtain
\begin{equation}
\frac{|V_{cd}|f_+^{(D\to\pi)}(0)}{|V_{cs}|f_+^{(D\to K)}(0)} = 0.188 \pm 0.008 \pm 0.002.  
\end{equation}
We can combine this result with calculations of $f(0)^{(D\to\pi)}/f(0)^{(D\to K)}$ to obtain
the ratio of CKM elements.  A recent light cone sum rules (LCSR) calculation, for example,
obtains \cite{Ball:2006yd} $f_+^{(D\to\pi)}(0)/f_+^{(D\to K)}(0) = 0.84\pm0.04$, which implies
\begin{equation}
\frac{|V_{cd}|}{|V_{cs}|} = 0.223 \pm 0.010_{\text{stat}} \pm 0.003_{\text{syst}} \pm 0.011_{\text{LCSR}}.  
\end{equation}

\section{Summary}
\label{sec:summary}
In summary, we have measured branching fractions and 
branching-fraction ratios for four semileptonic $D$ decay modes in five $q^2$
bins. The branching fraction results are the most precise measured to date and
agree well with world averages~\cite{pdg}.  
Our modified pole $\alpha$ parameter results agree within 1.3$\sigma$
with previous determinations by CLEO III
\cite{cleo_2005}, FOCUS \cite{focus_2005}, and $Ke\nu$
results from Belle\cite{belle_2006}, but show over $3\sigma$ disagreement
with Belle $K\mu\nu$ results and LQCD fits.  The $\alpha$ parameters
obtained with our individual $Ke\nu$ results are separated from the
recent BaBar result \cite{Aubert:2006mc} by about $2.5\sigma$.  
The $z$ expansion results between BaBar and our $Ke\nu$ agree closer to the $2\sigma$ level or better, depending on the level of correlation between the BaBar $r_1$ and $r_2$ parameters.  
The discrepancy with LQCD is difficult to quantify because the
covariance matrix for the LQCD form factors is lost during
the chiral extrapolation procedure for the published analysis \cite{fnalqcd}.
We have made the most precise CKM
determinations from $D$ semileptonic decays to date, and the results
agree very well with neutrino based determinations of $|V_{cd}|$ and
charmed-tagged $W$ decay measurements of $|V_{cs}|$ \cite{pdg}.  Overall, 
these measurements represent a marked improvement in our knowledge
of $D$ semileptonic decay.

We gratefully acknowledge the effort of the CESR staff 
in providing us with excellent luminosity and running conditions. 
D.~Cronin-Hennessy and A.~Ryd thank the A.P.~Sloan Foundation. 
This work was supported by the National Science Foundation,
the U.S. Department of Energy, and 
the Natural Sciences and Engineering Research Council of Canada.

\appendix
\section{The $z$-expansion: detailed forms and alternate results}
\label{app:zexpand}
\begin{table*}[t]
\caption{\label{tab:alternateT0} The fit results for the 3 parameter $z$--expansion fit for both $t_0=0$ and 
$t_0=t_+\left(1-\sqrt{1-t_-/t_+}\right)$.  The fit results are also presented in terms of the ratios
$r_i = a_i / a_0$ for $i=1,2$.}
\begin{ruledtabular}
\begin{tabular}{ccllllll}
 & &            $a_0$       & $a_1$        & $a_2$         & $\rho_{01}$ & $\rho_{02}$ & $\rho_{12}$ \\
$t_0 = 0$   & $\pi^- e^+ \nu_e$   & 0.044(2)(1) & -0.18(7)(2)  &  -0.03(35)(12)& 0.81        & 0.71        & 0.96\\
            & $\pi^0 e^+ \nu_e$   & 0.044(3)(1) & -0.23(11)(2) & -0.60(57)(15) & 0.80        & 0.67        & 0.95 \\
            & $K^- e^+ \nu_e$     & 0.0234(3)(3)& -0.009(21)(7)& 0.52(28)(6)   & 0.62        & 0.56        & 0.96  \\
            & $\bar{K}^0e^+\nu_e$ & 0.0224(4)(3)& 0.009(32)(7) & 0.76(42)(8)   & 0.72        & 0.64        & 0.96 \\
            &                   & $a_0$       & $r_1$        & $r_2$         & $\rho_{01}$ & $\rho_{02}$ & $\rho_{12}$ \\
$t_0 = 0$   & $\pi^- e^+ \nu_e$   & 0.044(2)(1) & -4.1(1.7)(0.6)&-0.7(8.0)(2.9)& 0.85        & 0.71        & 0.95\\
            & $\pi^0 e^+ \nu_e$   & 0.044(3)(1) & -5.3(2.8)(0.5)& -14(14)(3)   & 0.85        & 0.71        & 0.95 \\
            & $K^- e^+ \nu_e$     & 0.0234(3)(3)& -0.4(9)(3)    & 22(12)(2)    & 0.62        & 0.54        & 0.95  \\
            & $\bar{K}^0e^+\nu_e$ & 0.0224(4)(3)& 0.4(1.4)(3)   & 34(18)(4)    & 0.72        & 0.61        & 0.96 \\
            &                   & $a_0$       & $a_1$        & $a_2$         & $\rho_{01}$ & $\rho_{02}$ & $\rho_{12}$ \\
$t_0=\tmin$ & $\pi^- e^+ \nu_e$   & 0.072(2)(1) & -0.15(5)(2)  &  -0.09(35)(13)& -0.48        & 0.21        & -0.94\\
            & $\pi^0 e^+ \nu_e$   & 0.065(4)(1) & -0.01(10)(2) & -0.63(57)(14) & -0.65        & 0.41        & -0.95 \\
            & $K^- e^+ \nu_e$     & 0.0252(2)(3)& -0.062(10)(2)& 0.52(28)(6)   & -0.14        &-0.24        & -0.79  \\
            & $\bar{K}^0e^+\nu_e$ & 0.0239(3)(3)& -0.067(15)(4) & 0.76(42)(9)  & -0.08        &-0.28        & -0.82 \\
            &                   & $a_0$       & $r_1$        & $r_2$         & $\rho_{01}$ & $\rho_{02}$ & $\rho_{12}$ \\
$t_0=\tmin$ & $\pi^- e^+ \nu_e$   & 0.072(2)(1) & -2.1(7)(3)   &  -1.2(4.8)(1.7)& -0.41        & 0.22        & -0.96\\
            & $\pi^0 e^+ \nu_e$   & 0.065(4)(1) & -0.2(1.5)(4) & -9.8(9.1)(2.1) & -0.64        & 0.47        & -0.97 \\
            & $K^- e^+ \nu_e$     & 0.0252(2)(3)& -2.4(4)(1)   & 21(11)(2)      & -0.05        & -0.27       & -0.81  \\
            & $\bar{K}^0e^+\nu_e$ & 0.0239(3)(3)& -2.8(6)(2)   & 32(18)(4)      &  0.004        & -0.31        & -0.84 \\
\end{tabular}
\end{ruledtabular}
\end{table*}

The standard choice for the outer function $\phi(t,t_0)$ in the $z$ expansion for $f_+(q^2)$ (Eq.~\ref{eq:zexpand})
arises from considerations of unitarity.  From a perturbative Operator Product Expansion (OPE) calculation, one can
show \cite{boyd_95,boyd_97,Bourrely:1980gp} that the choice
\begin{eqnarray}
\phi(t,t_0) & = & \alpha 
\left(\sqrt{t_+ - t}+\sqrt{t_+ - t_0}\right) \\
& \times & \frac{t_+ - t}{(t_+ - t_0)^{1/4}} 
\frac{(\sqrt{t_+ - t}+\sqrt{t_+ - t_-})^{3/2}}{(\sqrt{t_+ - t}+\sqrt{t_+})^5} \nonumber
\end{eqnarray}
leads to a constraint on the coefficients
\begin{equation}
\sum_{k=0}^{n_a}a_k^2 \leq 1,
\end{equation}
for any choice of $n_a$.  The bound corresponds to forbidding the production rate of $D\pi$ states by
the relevant current to exceed the inclusive production rate, which can be calculated within the OPE.
To leading order, the coefficient $\alpha$ is given by
\begin{equation}
\alpha = \sqrt{\frac{\pi m_c^2}{3}}.
\end{equation}
Numerically, we have taken the charm quark mass to be $m_c=1.2$ GeV.

The choice of the parameter $t_0$ within the $z$--expansion provides a potential source of ambiguity
when comparing experimental results.
In our fits, we have, for simplicity, chosen  $t_0=0$ in our form factor fits utilizing the $z$--expansion.  
Another common choice for $t_0$ is that which minimizes the maximal value of the mapping $z(q^2)$ over the
entire physical range.  The  value $t_0=t_+\left(1-\sqrt{1-t_-/t_+}\right)$, where $t_\pm=m_D\pm m_{K,\pi}$
accomplishes this minimization.  The best fit $a_i$ values for our three parameter fit using this alternate value for 
$t_0$ are presented in Table~\ref{tab:alternateT0}.  The values for $|V_{cq}|f_+(0)$ ($q=s,d$) that we
find in these fits are identical, within the precision we are quoting, to those presented in Table~\ref{table:all_ff}.

Finally, some experimental results for the $z$--expansion are presented in terms of the 
ratios $r_i = a_i / a_0$ for $i>0$.
To allow straightforward comparison, we also quote our results in this form in Table~\ref{tab:alternateT0}.

\begin{turnpage}
\begingroup
\squeezetable
\begin{table*}
\caption{\label{tab:statCorrel} The statistical correlation matrix obtained from the simultaneous fit to the data (see Section~\protect\ref{sec:fit_comp}).  The lines indicate the mode boundaries.  The modes are labeled by their final state hadron. Within each submode, the five $q^2$ intervals are ordered from lowest  to highest.}
\begin{ruledtabular}
\begin{tabular}{lddddd|ddddd|ddddd|ddddd}
&\multicolumn{5}{c|}{$\pi^-$}&\multicolumn{5}{c|}{$\pi^0$}&\multicolumn{5}{c|}{$K^-$}&\multicolumn{5}{c}{$\bar{K}^0$} \\ 
             & 1.000 &-0.047 & 0.034 & 0.025 & 0.030 &-0.002 & 0.002 & 0.003 & 0.004 & 0.002 &-0.059 & 0.003 & 0.002 & 0.002 & 0.000 &-0.010 & 0.006 & 0.009 & 0.010 & 0.006 \\
             &       & 1.000 &-0.045 & 0.034 & 0.035 & 0.001 &-0.005 & 0.003 & 0.006 & 0.004 &-0.007 &-0.026 & 0.004 & 0.003 & 0.000 & 0.002 &-0.014 & 0.011 & 0.014 & 0.009 \\
$\pi^-$&       &       & 1.000 &-0.044 & 0.034 & 0.001 & 0.002 &-0.008 & 0.004 & 0.006 & 0.000 &-0.009 &-0.013 & 0.001 & 0.000 & 0.004 & 0.003 &-0.019 & 0.011 & 0.010 \\
             &       &       &       & 1.000 &-0.016 & 0.001 & 0.002 & 0.004 &-0.022 & 0.006 &-0.001 & 0.000 &-0.009 &-0.011 & 0.001 & 0.002 & 0.007 &-0.003 &-0.038 &-0.001 \\
             &       &       &       &       & 1.000 &-0.001 &-0.001 &-0.001 & 0.007 &-0.115 &-0.002 &-0.002 &-0.003 &-0.021 &-0.030 &-0.004 &-0.002 &-0.004 &-0.022 &-0.053 \\ \hline 
             &       &       &       &       &       & 1.000 &-0.089 & 0.033 & 0.017 & 0.018 & 0.000 & 0.000 & 0.000 & 0.000 &-0.001 &-0.013 & 0.006 & 0.005 & 0.004 & 0.001 \\
             &       &       &       &       &       &       & 1.000 &-0.094 & 0.032 & 0.023 & 0.001 &-0.001 & 0.000 & 0.001 & 0.000 &-0.006 &-0.010 & 0.006 & 0.004 & 0.001 \\
$\pi^0$&       &       &       &       &       &       &       & 1.000 &-0.090 & 0.032 & 0.001 & 0.001 &-0.002 & 0.001 & 0.000 & 0.004 &-0.006 &-0.016 & 0.002 & 0.002 \\
             &       &       &       &       &       &       &       &       & 1.000 &-0.069 & 0.001 & 0.001 & 0.000 &-0.004 &-0.002 & 0.002 & 0.004 &-0.013 &-0.029 &-0.005 \\
             &       &       &       &       &       &       &       &       &       & 1.000 & 0.001 & 0.001 & 0.000 &-0.005 &-0.012 &-0.002 &-0.004 &-0.003 &-0.028 &-0.051 \\ \hline 
             &       &       &       &       &       &       &       &       &       &       & 1.000 &-0.064 & 0.023 & 0.017 & 0.012 &-0.033 & 0.006 & 0.005 & 0.005 & 0.002 \\
             &       &       &       &       &       &       &       &       &       &       &       & 1.000 &-0.070 & 0.021 & 0.011 & 0.002 &-0.019 & 0.006 & 0.007 & 0.003 \\
$K^-$&       &       &       &       &       &       &       &       &       &       &       &       & 1.000 &-0.071 & 0.013 & 0.007 & 0.005 &-0.021 & 0.001 & 0.002 \\
             &       &       &       &       &       &       &       &       &       &       &       &       &       & 1.000 &-0.094 & 0.005 & 0.007 & 0.000 &-0.040 &-0.016 \\
             &       &       &       &       &       &       &       &       &       &       &       &       &       &       & 1.000 & 0.000 & 0.000 &-0.001 &-0.019 &-0.062 \\ \hline 
             &       &       &       &       &       &       &       &       &       &       &       &       &       &       &       & 1.000 &-0.068 & 0.031 & 0.019 & 0.007 \\
             &       &       &       &       &       &       &       &       &       &       &       &       &       &       &       &       & 1.000 &-0.060 & 0.027 & 0.009 \\
$\bar{K}^0$&       &       &       &       &       &       &       &       &       &       &       &       &       &       &       &       &       & 1.000 &-0.068 & 0.011 \\
             &       &       &       &       &       &       &       &       &       &       &       &       &       &       &       &       &       &       & 1.000 &-0.098 \\
             &       &       &       &       &       &       &       &       &       &       &       &       &       &       &       &       &       &       &       & 1.000 \\
\end{tabular}
\end{ruledtabular}

\caption{\label{tab:systCorrel} The total systematic correlation matrix for the 20 measured mode /  $q^2$ intervals (see Section~\protect\ref{sec:syst_errors}).  The lines indicate the mode boundaries. The modes are labeled by their final state hadron. Within each mode, the five $q^2$ intervals are ordered from lowest  to highest.}
\begin{ruledtabular}
\begin{tabular}{lddddd|ddddd|ddddd|ddddd}
&\multicolumn{5}{c|}{$\pi^-$}&\multicolumn{5}{c|}{$\pi^0$}&\multicolumn{5}{c|}{$K^-$}&\multicolumn{5}{c}{$\bar{K}^0$} \\ 
             & 1.00 & 0.73 & 0.64 & 0.59 & 0.51 & 0.45 & 0.29 & 0.13 & 0.28 & 0.06 & 0.75 & 0.59 & 0.56 & 0.53 & 0.59 & 0.34 & 0.27 & 0.22 & 0.26 & 0.26 \\
             &      & 1.00 & 0.87 & 0.81 & 0.56 & 0.13 & 0.33 & 0.21 & 0.34 &-0.04 & 0.75 & 0.74 & 0.74 & 0.70 & 0.63 & 0.28 & 0.32 & 0.32 & 0.26 & 0.13 \\
$\pi^-$&      &      & 1.00 & 0.94 & 0.46 & 0.23 & 0.55 & 0.34 & 0.39 &-0.20 & 0.79 & 0.86 & 0.86 & 0.85 & 0.64 & 0.25 & 0.32 & 0.28 & 0.19 &-0.10 \\
             &      &      &      & 1.00 & 0.46 & 0.19 & 0.61 & 0.38 & 0.44 &-0.21 & 0.77 & 0.89 & 0.89 & 0.91 & 0.65 & 0.23 & 0.30 & 0.27 & 0.17 &-0.11 \\
             &      &      &      &      & 1.00 & 0.05 & 0.14 & 0.08 & 0.10 & 0.24 & 0.57 & 0.52 & 0.52 & 0.50 & 0.56 & 0.20 & 0.18 & 0.19 & 0.22 & 0.37 \\ \hline 
             &      &      &      &      &      & 1.00 & 0.70 & 0.77 & 0.68 & 0.41 & 0.35 & 0.27 & 0.24 & 0.22 & 0.20 & 0.56 & 0.53 & 0.47 & 0.49 & 0.31 \\
             &      &      &      &      &      &      & 1.00 & 0.83 & 0.75 & 0.01 & 0.46 & 0.58 & 0.57 & 0.62 & 0.32 & 0.56 & 0.62 & 0.56 & 0.46 & 0.07 \\
$\pi^0$&      &      &      &      &      &      &      & 1.00 & 0.71 & 0.31 & 0.28 & 0.37 & 0.36 & 0.39 & 0.19 & 0.40 & 0.45 & 0.42 & 0.34 & 0.07 \\
             &      &      &      &      &      &      &      &      & 1.00 & 0.36 & 0.44 & 0.49 & 0.48 & 0.48 & 0.31 & 0.64 & 0.70 & 0.68 & 0.63 & 0.32 \\
             &      &      &      &      &      &      &      &      &      & 1.00 &-0.04 &-0.15 &-0.14 &-0.19 & 0.03 & 0.29 & 0.24 & 0.28 & 0.36 & 0.57 \\ \hline 
             &      &      &      &      &      &      &      &      &      &      & 1.00 & 0.93 & 0.92 & 0.86 & 0.75 & 0.40 & 0.39 & 0.35 & 0.30 & 0.11 \\
             &      &      &      &      &      &      &      &      &      &      &      & 1.00 & 1.00 & 0.98 & 0.81 & 0.33 & 0.40 & 0.38 & 0.31 & 0.03 \\
$K^-$  &      &      &      &      &      &      &      &      &      &      &      &      & 1.00 & 0.98 & 0.82 & 0.33 & 0.40 & 0.38 & 0.31 & 0.05 \\
             &      &      &      &      &      &      &      &      &      &      &      &      &      & 1.00 & 0.82 & 0.28 & 0.37 & 0.35 & 0.28 & 0.02 \\
             &      &      &      &      &      &      &      &      &      &      &      &      &      &      & 1.00 & 0.19 & 0.30 & 0.30 & 0.38 & 0.37 \\ \hline 
             &      &      &      &      &      &      &      &      &      &      &      &      &      &      &      & 1.00 & 0.96 & 0.94 & 0.88 & 0.62 \\
             &      &      &      &      &      &      &      &      &      &      &      &      &      &      &      &      & 1.00 & 0.99 & 0.94 & 0.63 \\
$\bar{K}^0$  &      &      &      &      &      &      &      &      &      &      &      &      &      &      &      &      &      & 1.00 & 0.96 & 0.67 \\
             &      &      &      &      &      &      &      &      &      &      &      &      &      &      &      &      &      &      & 1.00 & 0.80 \\
             &      &      &      &      &      &      &      &      &      &      &      &      &      &      &      &      &      &      &      & 1.00 \\
\end{tabular}
\end{ruledtabular}
\end{table*}
\endgroup
\end{turnpage}

\section{Correlation Matrices}
\label{correl}
To allow complete external use of the partial branching fractions presented in this paper, we present the statistical and systematic uncertainty correlation matrices.  These matrices will allow, for example, for simultaneous fits of these results with other experimental results to obtain form factor parameters.
The statistical correlation matrix (Table~\ref{tab:statCorrel}) is derived from the $20\times 20$ covariance matrix produced in our fitting procedure.  

To obtain the systematic correlation matrix (Table~\ref{tab:systCorrel}), we create a separate covariance matrix from the correlated motions of all 20 yields in each individual systematic study.  We then sum the resulting matrices to obtain the total systematic covariance matrix.  In the absence of correlations, this procedure would reduce to adding the systematic contributions for a given measurement in quadrature.
In producing the covariance matrix for the form factor systematic uncertainty, we assume that the two pion modes are fully correlated and similarly for the two kaon modes, but treat the pion and kaon uncertainties as uncorrelated.  For the $N_{D^+D^-}$ and $N_{D^0\bar{D}^0}$ uncertainties, we take into account the 39\% correlation  in those yields.

\end{document}